\documentclass[aps,prb,twocolumn,superscriptaddress,showpacs,floatfix]{revtex4}

\usepackage{amsfonts,amssymb,amsmath,bbm,graphicx}

\newcommand{\eqn}[1]{\begin{equation} #1 \end{equation}} 
\newcommand{\aln}[1]{\begin{align} #1 \end{align}}       
\newcommand{\mul}[1]{\begin{multline} #1 \end{multline}} 
\newcommand{\mc}{\mathcal}                               
\newcommand{\msf}{\mathsf}                               
\newcommand{\1}{\mathbbm 1}              
\newcommand{\bs}{\boldsymbol}                  
\newcommand{\eq}[1]{(\ref{#1})}                
\newcommand{\pd}{\partial}                     
\newcommand{\wtilde}{\widetilde}               
\newcommand{\wbar}{\overline}                  

\begin{document}


\title{Semiclassical theory of weak antilocalization and spin
relaxation\\ in ballistic quantum dots}


\author{Oleg Zaitsev}
\email[E-mail: ]{oleg.zaitsev@physik.uni-regensburg.de}
\affiliation{Institut f\"ur Theoretische Physik, Universit\"at Regensburg,
D-93040 Regensburg, Germany}

\author{Diego Frustaglia}
\affiliation{NEST-INFM \& Scuola Normale Superiore, 56126 Pisa, Italy}

\author{Klaus Richter}
\affiliation{Institut f\"ur Theoretische Physik, Universit\"at Regensburg,
D-93040 Regensburg, Germany}



\begin{abstract}

We develop a semiclassical theory for spin-dependent quantum transport in
ballistic quantum dots. The theory is based on the semiclassical Landauer
formula, that we generalize to include spin-orbit and Zeeman interaction.
Within this approach, the orbital degrees of freedom are treated
semiclassically, while the spin dynamics is computed quantum mechanically.
Employing this method, we calculate the quantum correction to the conductance
in quantum dots with Rashba and Dresselhaus spin-orbit interaction. We find a
strong sensitivity of the quantum correction to the underlying classical
dynamics of the system. In particular, a suppression of weak antilocalization
in integrable systems is observed. These results are attributed to the
qualitatively different types of spin relaxation in integrable and chaotic
quantum cavities. 

\end{abstract}


\maketitle



\section{Introduction}

Guided by the vision to incorporate spin physics into the far-advanced
semiconductor (hetero)structure technology, semiconductor-based spin
electronics (see, e.g., Ref.~\onlinecite{zuti04}) has developed into a
prominent branch of present spintronics research. In this context spin-orbit
(SO) interactions have recently received considerable attention since they give
rise to rich spin dynamics and a variety of spin phenomena in nonmagnetic
semiconductors. Though SO couplings have been a subject of continuous research
throughout the last decades~\cite{rash60, bych84, dres55, dyak71, dyak86,
lomm88, andr94}, there is presently a revival in investigating SO effects owing
to their role in spin transistors~\cite{datt90, mire01}, spin
interferometers~\cite{nitt99, frus04a}, spin filters~\cite{gove02b, kise01},
and spin pumps~\cite{gove03, shar03}, to name only a few examples. Furthermore,
most recently the intrinsic spin Hall effect~\cite{sino04} in a SO-coupled
system has caused an intense and controversial discussion in the
literature~\cite{zhan05}. Finally, in spin-based quantum computation SO-induced
spin relaxation effects may play a role~\cite{golo04}.

The interplay between spin dynamics and confinement effects is particularly
intriguing in quantum transport through low-dimensional devices at low
temperatures where quantum coherence effects additionally arise. There exist
two prominent experimental probes for SO effects in quantum transport:  (i) 
characteristic beating patterns in Shubnikov-de Haas oscillations in
two-dimensional electron gases with tunable SO coupling, controlled via a
back-gate voltage~\cite{luo88, nitt97, enge97, scha04}, and (ii) weak
antilocalization~\cite{schi02, mill03, meij04} (WA), an enhancement of the
magnetoconductance at zero magnetic field  owing to spin-dependent quantum
interference effects. Since systems without SO coupling exhibit weak
localization (WL), i.e., a reduction in the magnetoconductance, the appearance
of WA allows conclusions to be drawn on the SO strength. While WA is fairly
well understood for disordered bulk systems~\cite{hika80, berg84, chak86}, in
recent experiments using ballistic bismuth~\cite{hack03} and GaAs~\cite{zumb02}
cavities, WA has been employed to study SO-induced spin dynamics and spin
relaxation phenomena in confined systems. These measurements are focussed on
the interesting inter-relation between quantum confined orbital motion and spin
evolution and relaxation in clean ballistic quantum dots. In these systems, 
the elastic mean free path is exceeding their size, and impurity scattering is
replaced by reflections off the system boundaries.

Corresponding efforts in treating SO~effects on spectra~\cite{vosk01, gove02a,
bulg03, saic02}, spin relaxation~\cite{khae00, wood02, pare02}, and the
interplay between SO and Zeeman coupling~\cite{brou02, falk02} in quantum dots
have also been made on the theoretical side. SO-induced WA in ballistic quantum
dots has been studied using random-matrix theory~\cite{alei01,crem03} (RMT) and
semiclassical approaches~\cite{zait04, zait05}. While RMT approaches are
restricted to quantum dots with corresponding chaotic classical dynamics, the
semiclassical transport theory comprises a much broader class of systems,
including integrable confinement geometries. Related semiclassical techniques
have also been applied to spin transmission~\cite{chan04a} and spin
relaxation~\cite{chan04b} in quantum dots. 

The purpose of the present article is a detailed exposition and extension of
the semiclassical methods of Refs.~\onlinecite{zait04} and~\onlinecite{zait05}.
The theory to be discussed here unifies two subject areas: the semiclassical
description of WL~\cite{bara93a, bara93b, rich02} and the semiclassical
treatment of SO~interaction~\cite{litt92, fris93, aman02, plet02, zait02,
plet03, cser04}.  Compared to the earlier works~\cite{zait04, zait05, chan04b}
on the subject, here we give special attention to the differences in spin
relaxation along open and closed trajectories, analyze the interplay between
Rashba and Dresselhaus interaction, and report on the full quantum calculations
of spin-dependent transmission and reflection.

This paper is organized as follows: In the introductory Sec.~\ref{sost}, using
path integrals with spin coherent states we deduce a spin-dependent
semiclassical propagator and the corresponding Green function. Our main
analytical results are presented in Sec.~\ref{semland}. There, on the basis of
Green functions a semiclassical approximation to the Landauer formula with spin
is derived. The semiclassical Landauer formula is then applied to chaotic
quantum dots, whereby the quantum corrections to transmission and reflection
are calculated. In Sec.~\ref{sr} we discuss how WA is related to the spin
evolution. We define measures for spin relaxation and consider, as an example, 
the spin relaxation in diffusive systems. In the following two sections the
general theory is applied to chaotic and integrable quantum dots with Rashba
and Dresselhaus SO~interaction. In Sec.~\ref{rdso} a detailed numerical study
of the spin relaxation is followed by an analysis of the limit of slow spin
dynamics (i.e., extremely weak SO~coupling). Additionally, we examine a gauge
transformation of the spin-orbit Hamiltonian that can be carried out in this
limit. The dependence of the quantum corrections to transmission and reflection
on the SO-coupling strength and magnetic field (Aharonov-Bohm and Zeeman
contributions) is presented in Sec.~\ref{rdqc}. There some of the semiclassical
numerical results are compared with full-scale numerical quantum calculations.

\section{Spin-orbit interaction in a semiclassical theory}
\label{sost}

In this preparatory section we construct a spin-dependent semiclassical
propagator\footnote{Another derivation involving a slightly different approach
can be found in Ref.~\onlinecite{chak86}.} and related Green function. It fully
describes the system at a given level of approximation and, thus, can serve as
a starting point for our derivation of a semiclassical Landauer formula for
systems with SO and Zeeman interaction (Sec.~\ref{semland}). 

In the spinless case, the semiclassical propagator is conventionally obtained
from the path-integral representation of the exact propagator~\cite{berr72}.
After the stationary-phase evaluation, which is valid in the semiclassical
limit, the classical trajectories are selected from all the paths in the
integral. In order to include spin into the path integral, a continuous basis
of spin states is required. The \emph{spin coherent states} represent such a
basis~\cite{klau79, koch95}. 

Following Ref.~\onlinecite{koch95}, we define a coherent state of spin $s =
\frac 1 2, 1, \ldots$ by 
\eqn{
  |\zeta \rangle = (1 + |\zeta|^2)^{-s} \exp\, (\zeta \hat s_+)\: 
  | \sigma = -s \rangle,
\label{scs}
}
where $\zeta$ is a complex number that labels the state, $\hat s_+ = \hat s_x +
i \hat s_y$ is the spin operator, and $|\sigma \rangle$ are the eigenstates
of~$\hat s_z$ with eigenvalues $\sigma = -s, \ldots, s$. To each $\zeta$
corresponds a three-dimensional unit vector $\mathbf n (\zeta) = \langle \zeta
| \hat{\mathbf s} | \zeta \rangle / s$ that denotes the spin direction. It is
easy to show that $\zeta$ is a stereographic projection from  the unit sphere
centered at the origin onto the plane $z=0$. The projection is given by $(n_x,
n_y, n_z) \longmapsto (\text{Re}\, \zeta, -\text{Im}\, \zeta, 0)$, where, in
particular, the south pole is mapped to $\zeta = 0$. In general, coherent
states have the minimal uncertainty of $\hat{\mathbf s}$ among all spin states
and are characterized by three real parameters: the direction $\mathbf n$ and
an overall phase. (Hence, any state of spin~$1/2$ is coherent.) In the current
definition, the phase is assigned to each $\mathbf n$ by Eq.~\eq{scs}, but
other phase assignments are possible.  Note that the phase of the state $|
\zeta = \infty \rangle \propto |\sigma = s \rangle$ with $\mathbf n = (0, 0,
1)$ is not well defined. However, this manifestation of the fundamental problem
of phase assignment~\cite{wu75} does not pose a difficulty in our case, since
the final results will be transformed to the $|\sigma \rangle$~representation
using the projection operators~$|\zeta \rangle \langle \zeta |$.  The states
\eq{scs} are normalized to unity, but, obviously, not mutually orthogonal (no
more than $2s + 1$ states of spin $s$ can be mutually orthogonal).
Nevertheless, having the property of resolution of unity,
\eqn{
  \int |\zeta \rangle \langle \zeta |\, d \mu (\zeta) = \hat \1, \quad 
  d \mu (\zeta) = \frac {2s + 1} {\pi\, (1 + |\zeta|^2)^2} d^2 \zeta ,
\label{ru}
}
they form an (overcomplete) basis and enable a path-integral construction.

Let us consider a rather general case of a system with Hamiltonian linear in
the spin operator~$\hat{\mathbf s}$:
\eqn{
  \hat H =  \hat H^0 (\hat {\mathbf q}, \hat {\mathbf p}) + \hbar \hat{\mathbf
  s} \cdot \hat {\mathbf C} (\hat {\mathbf q}, \hat {\mathbf p}).
\label{soham}
}
Here $\hat {\mathbf q}$ and $\hat {\mathbf p}$ are the $d$-dimensional
coordinate and momentum operators, respectively, $\hat H^0 (\hat {\mathbf q},
\hat {\mathbf p})$ is the spin-independent Hamiltonian, and $\hbar \hat{\mathbf
s} \cdot\hat {\mathbf C} (\hat {\mathbf q}, \hat {\mathbf p})$ describes the
SO interaction and the Zeeman interaction with an external (generally
inhomogeneous) magnetic field. Utilizing the idea of Refs.~\onlinecite{plet02}
and~\onlinecite{plet03}, we express the propagator in the combined coordinate
and spin-coherent-state representation in terms of the path integral
\aln{
  &U(\mathbf q_2, \zeta_2, \mathbf q_1, \zeta_1; T) \equiv 
  \langle \mathbf q_2, \zeta_2 | e^{ -(i/ \hbar) \hat H T} |
  \mathbf q_1, \zeta_1 \rangle \notag \\  
  &= \int \frac {\mc D [\mathbf q] \mc D [\mathbf p]} {(2 \pi \hbar)^d} \mc D
  \mu [\zeta] \, \exp \left\{\frac i \hbar \mc W[\mathbf q, \mathbf p, \zeta;
  T]  \right\}.
\label{pi}
}
The integration is performed over the paths $[\mathbf q (t), \mathbf p (t),
\zeta (t)]$ in the spin-orbit phase space connecting $(\mathbf q_1, \mathbf
p_1, \zeta_1)$ to $(\mathbf q_2, \mathbf p_2, \zeta_2)$ in time~$T$ with
arbitrary $\mathbf p_1$ and~$\mathbf p_2$. The integration measures are defined
by 
\eqn{
  \frac {\mc D [\mathbf q] \mc D [\mathbf p]} {(2 \pi \hbar)^d} \mc D
  \mu [\zeta] =
  \underset{n \to \infty} {\text{lim}} \prod_{j=1}^{n-1} \frac {d \mathbf q
  (t_j)\, d \mathbf p (t_j)} {(2 \pi \hbar)^d}\, d \mu \bigl(\zeta(t_j) \bigr),
}
where $t_j = j T/ n$. The Hamilton principal function $\mc W = \mc W^0 +
\hbar s \mc W^1$ consists of two contributions: the usual classical part,
\eqn{
  \mc W^0 [\mathbf q, \mathbf p; T] = \int_0^T dt\, [\mathbf p \cdot
  \dot{\mathbf q} - H^0 (\mathbf q, \mathbf p)],
}
and the spin-related part,
\eqn{
  \mc W^1[\mathbf q, \mathbf p, \zeta; T] = \int_0^T dt \left[  \frac {\zeta
  \dot \zeta^* - \zeta^* \dot \zeta} {i (1 + |\zeta|^2)} - \mathbf n (\zeta)
  \cdot \mathbf C (\mathbf q, \mathbf p) \right].
}
Now we can separate the integration over the spin paths in~\eq{pi}, thereby
representing the propagator~as~\cite{plet03}
\mul{
  U(\mathbf q_2, \zeta_2, \mathbf q_1, \zeta_1; T) = \\
  \int \frac {\mc D [\mathbf q] \mc D [\mathbf p]} {(2 \pi \hbar)^d} \,
  K_{[\mathbf q, \mathbf p]} (\zeta_2, \zeta_1; T) \,
  \exp \left\{\frac i \hbar \mc W^0[\mathbf q, \mathbf p; T] \right\}
\label{pisep}
}
with
\eqn{
  K_{[\mathbf q, \mathbf p]} (\zeta_2, \zeta_1; T) = \int \mc D \mu [\zeta] \,
  \exp \left\{i s \mc W^1[\mathbf q, \mathbf p, \zeta; T] \right\}.
\label{spinpi} 
}
Clearly, $K_{[\mathbf q, \mathbf p]} (\zeta_2, \zeta_1; T)$ is a propagator of
a system with the time-dependent Hamiltonian $\hat H_{[\mathbf q, \mathbf p]}
(t) = \hbar \hat{\mathbf s} \cdot \mathbf C_{[\mathbf q, \mathbf p]} (t)$,
where $\mathbf C_{[\mathbf q, \mathbf p]} (t) = \mathbf C \bigl(\mathbf q (t),
\mathbf p(t) \bigr)$ is calculated along the path $[\mathbf q (t), \mathbf
p(t)]$ of the integral~\eq{pisep}. This Hamiltonian describes a spin,
precessing in the time-dependent magnetic field $\mathbf C_{[\mathbf q, \mathbf
p]} (t)$.\footnote{$\mathbf C_{[\mathbf q, \mathbf p]} (t)$ is measured in
units of frequency.} Expression~\eq{spinpi} for $K_{[\mathbf q, \mathbf p]}
(\zeta_2, \zeta_1; T)$  can be integrated explicitly~\cite{koch95}, yielding
the usual spin propagator in the basis of coherent states (Appendix~\ref{sp}).

We proceed by evaluating the path integral~\eq{pisep} in the semiclassical
limit $\mc W^0 \gg \hbar$.  The integration simplifies considerably, if
the spin-dependent Hamiltonian is treated as a perturbation, i.e., when
\eqn{
  \hbar s |\mathbf C (\mathbf q, \mathbf p)| \ll |H^0 (\mathbf q, \mathbf p)|.
\label{wcl}
}
This condition, assumed for the rest of the paper, is usually fulfilled in
experiments based on the semiconductor heterostructures. According to this
requirement, the spin-precession length must be much larger than the Fermi
wavelength, however it can be smaller, of order, or greater than the system
size.  The semiclassical and perturbative regimes can be implemented
simultaneously~\cite{bolt98, bolt99} by formally letting $\hbar \to 0$ and
keeping all other quantities fixed. Then the phase of the integrand in
Eq.~\eq{pisep} is a rapidly varying functional, which justifies the use of the
stationary-phase approximation. It is crucial that $K_{[\mathbf q, \mathbf p]}
(\zeta_2, \zeta_1; T)$ does not depend on~$\hbar$, i.e., it is a slowly varying
functional, and, therefore, its effect on the stationary trajectories can be
neglected. Thus, the stationary trajectories are the extremals solely of $\mc
W^0[\mathbf q, \mathbf p; T]$, which  means that they are the classical orbits
of the \emph{spinless} Hamiltonian~$H^0$. The resulting semiclassical
propagator,
\mul{
  U_{\text{sc}} (\mathbf q_2, \zeta_2, \mathbf q_1, \zeta_1; T) \\
  = \sum_\gamma K_\gamma (\zeta_2, \zeta_1; T)\: \mc C_\gamma \exp \left\{\frac
  i \hbar \mc W^0_\gamma (\mathbf q_2, \mathbf q_1; T) \right\},
}
is a sum over all classical trajectories $\gamma \equiv [\mathbf q_\gamma (t),
\mathbf p_\gamma (t)]$ of time~$T$ from $\mathbf q_1$ to~$\mathbf q_2$. The
prefactor~$\mc C_\gamma$, arising from the stationary-phase integration, is the
same as in the spinless case~\cite{berr72}:
\eqn{
  \mc C_\gamma = \frac {\exp \left(-i \frac \pi 4 d -i \frac \pi 2 \nu_\gamma
  \right)} {(2 \pi \hbar)^{d/2}} 
  \left| \sideset{}{_{\alpha \beta}}\det \frac {\pd^2 \mc W^0_\gamma
  (\mathbf q_2, \mathbf q_1; T)} {\pd q_2^\alpha \pd q_1^\beta} \right|^{1/2},
}
where $\nu_\gamma$ is the Maslov index. Although the classical trajectories are
not affected by the spin motion, the reverse is not true. Indeed, the spin
propagator~$K_\gamma$, computed along the classical trajectories, describes the
spin evolution in the effective magnetic field $\mathbf C_\gamma (t) = \mathbf
C \bigl(\mathbf q_\gamma (t), \mathbf p_\gamma (t)\bigr)$ generated by these
trajectories.

The semiclassical Green function is given by the Laplace transform of
$U_{\text{sc}} (T)$ to the energy domain~$E$,
\eqn{
  G(E) = - \frac i \hbar \int_0^\infty dT e^{ i  (E + i0^+) T/ \hbar} \,
  U_{\text{sc}} (T),
}
evaluated in the stationary-phase approximation. As before, $K_\gamma (T)$ does
not modify the stationary-phase condition, and the theory without spin can be
applied. Moreover, using the resolution of unity~\eq{ru}, the spin propagator
can be transformed to the usual $|\sigma \rangle$~basis. Finally, we obtain
\mul{
  G_{\sigma^\prime \sigma} (\mathbf q_2, \mathbf q_1; E)  \\
  = \sum_\gamma \, (\hat K_\gamma)_{\sigma^\prime \sigma}   \, \mc F_\gamma 
  \exp \left\{\frac i \hbar \mc S^0_\gamma  (\mathbf q_2, \mathbf q_1; E)
  \right\}  
\label{semGF}
}
with $\sigma, \sigma^\prime = -s, \ldots, s$. In Eq.~\eq{semGF}, $\gamma$~is a
classical trajectory of the Hamiltonian $H^0 = E$ with the action $\mc
S^0_\gamma = \int_\gamma \mathbf p \cdot d \mathbf q$ and time $T_\gamma (E) =
\pd \mc S^0_\gamma / \pd E$. $\hat K_\gamma (t)$~is the operator form of the
spin propagator [Appendix~\ref{sp}, Eqs.~\eq{spinprop} and~\eq{spprgen}], and
$(\hat K_\gamma)_{\sigma^\prime \sigma} \equiv \langle \sigma^\prime | \, \hat
K_\gamma\, \bigl(T_\gamma (E) \bigr)\, | \sigma \rangle$ is its matrix element.
The prefactor is given by 
\eqn{
  \mc F_\gamma = \mc C_\gamma \exp ^{ -i \, \frac \pi  4 \,
  \text{sgn} \left(d T_\gamma / dE \right) },
}
and $\mc C_\gamma$ is expressed in terms of the derivatives of $\mc S^0_\gamma
(\mathbf q_2, \mathbf q_1; E)$ (Ref.~\onlinecite{berr72}).

\section{Semiclassical Landauer formula with spin}
\label{semland}

The semiclassical Landauer formula with spin, derived below, is the main
analytical result of this paper. It forms the basis for the subsequent
semiclassical treatment of the spin-dependent transport in two-dimensional
systems. 

\subsection{Derivation of the formula}

We start from the standard (quantum) Landauer formula, that relates the
conductance $(e^2/h) \mc T$ of a sample with two ideal leads to its
transmission coefficient~$\mc T$ (Ref.~\onlinecite{fish81}).  Assuming that the
leads support $N$ and $N^\prime$ open channels (not counting the spin
degeneracy), respectively, the transmission can be expressed as the sum
\eqn{
  \mc T = \sum_{n=1}^{N^\prime} \sum_{m=1}^N \sum_{\sigma^\prime\!,\, \sigma =
  -s}^s | t_{n \sigma^\prime\!,\, m \sigma} |^2.
\label{qutran}
}
Here $t_{n \sigma^\prime\!,\, m \sigma}$ is the transmission amplitude between
the incoming channel~$|m \sigma \rangle$ (with spin projection~$\sigma$) and
the outgoing channel $|n \sigma^\prime \rangle$ belonging to different leads. 
We shall also consider the reflection coefficient
\eqn{
  \mc R = \sum_{n,m = 1}^N \sum_{\sigma^\prime\!,\, \sigma = -s}^s \left| r_{n
  \sigma^\prime\!,\, m \sigma} \right|^2,
\label{qurefl}
}
where the reflection amplitude $r_{n \sigma^\prime\!,\, m \sigma}$ is defined
for the channels of the same lead. The transmission and reflection satisfy the
normalization condition 
\eqn{
  \mc T + \mc R = (2s + 1)\, N 
\label{norm}
}
that follows from the unitarity of the scattering matrix. 

Consider, as a model for a (large) quantum dot, a two-dimensional cavity
(billiard) with hard-wall leads. The particle in the cavity is subjected to the
SO and Zeeman interaction of the form~\eq{soham}. Semiclassical expressions for
the transition amplitudes in the spinless case were derived in
Refs.~\onlinecite{bara93a} and~\onlinecite{bara93b} by projecting a
semiclassical Green function onto the lead eigenstates, while integrating over
the lead cross sections in the stationary-phase approximation. For a particle
with spin we implement this procedure using the semiclassical Green
function~\eq{semGF}. In the semiclassical limit of large action, $\mc
S^0_\gamma \gg \hbar$, the spin-propagator element~$(\hat
K_\gamma)_{\sigma^\prime \sigma}$ does not shift the stationary point. In the
resulting expression,
\eqn{
  t_{n \sigma^\prime, m \sigma} = \sum_{\gamma (\bar n, \bar m)} (\hat
  K_\gamma)_{\sigma^\prime \sigma}\, \mc A_\gamma \exp \left( \frac i
  \hbar \mc S_\gamma \right), 
\label{tsemsp}
}
the only effect of spin is to weight the contribution of each trajectory in the
sum with the respective matrix element of~$\hat K_\gamma$. In Eq.~\eq{tsemsp}
$\gamma (\bar n, \bar m)$~is any classical trajectory of energy~$E$ that enters
(exits) the cavity at a certain angle $\Theta_{\bar m}$ ($\Theta_{\bar n}$)
measured from the normal at a lead's cross
section.\footnote{Equation~\eq{tsemsp} is only valid if there are no families
of direct trajectories (without reflections) connecting the
leads~\cite{bara93b}.} The angles are determined by the transverse momentum in
the leads: $\sin \Theta_{\bar m} = \bar m \pi / k w$ and $\sin \Theta_{\bar n}
= \bar n \pi / k w^\prime$, where $k$ is the wavenumber and $w$ and $w^\prime$
are the widths of the entrance and exit leads. The action for a trajectory of
length $L_\gamma$ is $\mc S_\gamma = \hbar k L_\gamma$. The prefactor is
given~by
\mul{
  \mc A_\gamma = - \sqrt{\frac {\pi \hbar} {2 w w^\prime}}
  \frac {\text{sgn} (\bar n)\, \text{sgn} (\bar m)} {\left| \cos \Theta_{\bar
  n} \cos \Theta_{\bar m}\, M_{21}^\gamma \right|^{1/2}} \\   
  \times \exp \left[ \vphantom{\frac 1 2} i k \left(\sin \Theta_{\bar m}\,
  y - \sin \Theta_{\bar n}\, y^\prime \right) \right.\\ 
  \left. {} - i \frac \pi 2 \left(\mu_\gamma -
  \frac 1 2 \right) \right],
}
where $M_{21}^\gamma$ is an element of the stability matrix (as defined, e.g.,
in Ref.~\onlinecite{brac03}), $y$ ($y^\prime$) is the coordinate on the lead's
cross section at which the orbit $\gamma$ enters (exits) the cavity, and
$\mu_\gamma$ is the modified Morse index~\cite{bara93b}. Substituting the
sum~\eq{tsemsp} and the corresponding result for~$r_{n \sigma^\prime, m
\sigma}$ in Eqs.~\eq{qutran} and~\eq{qurefl} we obtain the semiclassical
approximation for the total transmission and reflection:
\mul{
  (\mc T, \mc R) = \\
  \sum_{nm} \sum_{\gamma (\bar n, \bar m)} \sum_{\gamma^\prime
  (\bar n, \bar m)} \mc M_{\gamma, \gamma^\prime} \mc A_\gamma \mc
  A^*_{\gamma^\prime} \exp \left[ \frac i\hbar \left(\mc S_\gamma - \mc
  S_{\gamma^\prime} \right) \right].
\label{spinLand}
}
Here in the case of transmission (reflection) the paths $\gamma$ and
$\gamma^\prime$ connect different leads (return to the same lead). In this
expression each orbital contribution is weighted with the spin \emph{modulation
factor} 
\eqn{
  \mc M_{\gamma, \gamma^\prime} \equiv \text{Tr}\, (\hat K_\gamma \hat
  K_{\gamma^\prime}^\dag),
\label{modfact}
}
where the trace is taken in spin space. Equations \eq{spinLand} and
\eq{modfact} generalize the semiclassical Landauer formula~\cite{bara93a,
bara93b} to the case of spin-dependent transport.

\subsection{Leading semiclassical contributions for a spinless particle}
\label{spinless}

In the semiclassical limit the phases in Eq.~\eq{spinLand} are rapidly varying
functions of energy, unless $\gamma$ and $\gamma^\prime$ have equal or nearly
equal actions. Therefore, if one calculates the  transmission and reflection
\emph{averaged} over a small energy window, most of the terms in the double sum
will vanish. In the following, we review the leading contributions for a
spinless system ($\mc M_{\gamma, \gamma^\prime} \equiv 1$) with time-reversal
symmetry:
\begin{trivlist}

\item (i) The \textbf{classical} part consists of the terms with $\gamma^\prime
= \gamma$ (Ref.~\onlinecite{bara91}). Their fast-varying phases cancel
(including the phase in the prefactor). For a classically ergodic (in
particular, \emph{chaotic}) system one finds~\cite{rich02}
\eqn{
  \mc T_{\text{cl}}^{(0)} = \frac {N N^\prime} {N + N^\prime}, \quad 
  \mc R_{\text{cl}}^{(0)} = \frac {N^2} {N + N^\prime}.
\label{TRcl}
}
(the superscript refers to zero spin and zero magnetic field). This result can
be obtained using the sum rule~\cite{rich02}
\eqn{
  \sum_{\gamma (\bar n, \bar m)} |\mc A_\gamma|^2\, \delta (L - L_\gamma)
  \simeq (N + N^\prime)^{-1} P_L(L).
\label{sumrule}
}
It implies that  the length~$L$ of the classical trajectories is distributed
according to 
\eqn{
  P_L(L) \simeq \frac 1 {L_{\text{esc}}} \exp \left(- \frac L
  {L_{\text{esc}}} \right),
\label{PL}
}
in other words, the probability for a particle to stay in an open chaotic
cavity decreases exponentially with time. The average escape length is given~by
\eqn{
  L_{\text{esc}} = \frac {\pi\, \msf A_c} {w + w^\prime} = \frac {k\, \msf A_c}
  {N + N^\prime},
\label{Lesc}
} 
where $\msf A_c$ is the area of the cavity. It is assumed that $L_{\text{esc}}
\gg L_b$, where 
\eqn{
  L_b  = \pi \msf A_c / \msf P_c
\label{Lb}
}
is the average distance between two consecutive bounces at the
boundaries~\cite{niel99} and $\msf P_c$ is the perimeter. In an arbitrary
billiard the last expression is true if the average is taken over the ensemble
of chords with random initial position and boundary component of the velocity.
In ergodic billiards the average can, alternatively, be calculated along almost
any trajectory. 

\item (ii) The \textbf{diagonal} quantum correction is defined for  reflection
only. It contains the terms with $n=m$ and $\gamma^\prime = \gamma^{-1}$, where
$\gamma^{-1}$ is the time reversal of $\gamma$
(Ref.~\onlinecite{bara93b}).\footnote{In Ref.~\onlinecite{rich02} the diagonal
contribution is defined differently and includes the classical part.} Clearly,
when $n$ and $m$ are different channels, the orbits $\gamma (\bar n, \bar m)$
and $\gamma^\prime (\bar n, \bar m)$ cannot be the mutual time-reversals, since
reversing the time would exchange $\bar n$ and~$\bar m$. Again, the two actions
are equal, and the result for an ergodic system without spin
reads~\cite{rich02}
\eqn{
  \delta \mc R_{\text{diag}}^{(0)} = \frac N {N + N^\prime}.
\label{Rdiag}
}

\item (iii) The \textbf{loop} contribution consists of pairs of long orbits
that stay close to each other in the configuration space. One orbit of the pair
has a self-crossing with a small crossing angle $\varepsilon$, thus forming a
loop, its counterpart has an anticrossing.\footnote{A~generalized phase-space
approach for the loop contributions, also valid for hyperbolic systems with
smooth potential, is considered in Ref.~\onlinecite{ture03}.} Away from the
crossing region the orbits are located exponentially close to each other: they
are related by time reversal along the loop and coincide along the
tails~\cite{rich02,sieb01} (Fig.~\ref{looppair}). The action difference for
these orbits is of second order in~$\varepsilon$. For spinless chaotic systems
with hyperbolic dynamics the loop terms in~\eq{spinLand} yield~\cite{rich02}
\eqn{
  \delta \mc T_{\text{loop}}^{(0)} = - \frac {N N^\prime} {(N + N^\prime)^2}, 
  \quad 
  \delta \mc R_{\text{loop}}^{(0)} = - \frac {N (N + 1)} {(N + N^\prime)^2}.
\label{TRloop}
}

\begin{figure}[tbp]
  \includegraphics[width=.5 \linewidth, angle=0]{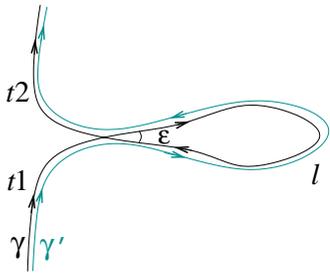}
  \caption{{\small Pair of orbits with a loop. (Neglecting the crossing region,
  we distinguish between the tails $t1, t2$ and the loop~$l$.)} }
\label{looppair}
\end{figure}

\end{trivlist}

From here~on, we will work in the limit $N,\, N^\prime \gg 1$. In this
semiclassical regime (in the leads)  the classical contribution~\eq{TRcl} (of
the order~$N$) is much greater than the quantum corrections \eq{Rdiag}
and~\eq{TRloop} (of the order~$N^0$), while higher-order loop corrections (of
order~$N^{-1}$ and smaller) can be neglected. We note that the normalization is
preserved order by order:
\aln{
  &\mc T_{\text{cl}}^{(0)}+ \mc R_{\text{cl}}^{(0)} = N, \\
  &\delta \mc R_{\text{diag}}^{(0)} + \delta \mc R_{\text{loop}}^{(0)} + \delta
  \mc T_{\text{loop}}^{(0)} = \mc O(N^{-1}).
\label{norm0}
}

\subsection{Spin-dependent quantum corrections to transmission and reflection}
\label{sdqc}

We now compute the spin modulation factor for the leading contributions to the
energy-smoothed $\mc T$ and $\mc R$, identified in Sec.~\ref{spinless}. First,
the case with time-reversal symmetry\footnote{This means that the orbital
Hamiltonian~$H^0$ has time-reversal symmetry and that $\mathbf C (\mathbf q,
\mathbf p )$ is an odd function of the velocity components, i.e., it changes
sign upon time reversal.} is considered:
\begin{trivlist}

\item (i) For the classical part we find, using the unitarity of~$\hat
K_\gamma$, that the modulation factor $\mc M_{\gamma, \gamma} = \text{Tr}\,
(\hat K_\gamma \hat K_\gamma^\dag) = 2s + 1$ reduces to the trivial spin
degeneracy. 

\item (ii) For the diagonal quantum correction the result is 
\eqn{
  \mc M_{\gamma, \gamma^{-1}} = \text{Tr}\, (\hat K_\gamma^2). 
\label{Mdiag}
}
It was taken into account that $\hat K_{\gamma^{-1}}^\dag = \hat K_\gamma$,
which follows from the relation $\mathbf C_{\gamma^{-1}} (t) = - \mathbf
C_\gamma (T_\gamma - t)$ and Eq.~\eq{spinprop}. 

\item (iii) Assuming that the trajectories forming a loop pair
(Fig.~\ref{looppair}) coincide along the tails $t1, t2$ and are mutually
time-reversed along the loop~$l$, thereby neglecting the crossing region, we
can represent the propagators as $\hat K_\gamma = \hat K_{t2} \hat K_l \hat
K_{t1}$ and $\hat K_{\gamma^\prime} = \hat K_{t2} \hat K_{l^{-1}} \hat K_{t1}$.
Hence, the modulation factor
\eqn{
  \mc M_{\gamma, \gamma^\prime} = \text{Tr}\, (\hat K_l^2). 
\label{Mloop}
}
is independent of the tails. 

\end{trivlist}

In the presence of a magnetic field the time-reversal symmetry is broken, and
the preceding results ought to be adjusted. In this paper we consider a
constant, uniform, arbitrarily directed magnetic field~$\mathbf B$. Its
component~$B_z$ normal to the cavity is assumed to be weak enough,\footnote{The
cyclotron radius must be larger than the system size.} so as not to change the
classical trajectories in~\eq{spinLand}, but only modify the action difference
by the Aharonov-Bohm (AB) phase. We define the AB~modulation factor
\eqn{
  \varphi_{\gamma, \gamma^\prime} \equiv \exp \left[ \frac i \hbar \Delta (\mc
  S_\gamma -  \mc S_{\gamma^\prime}) \right]
  = \exp \left( i \frac {4 \pi \msf A B_z} {\Phi_0} \right),
\label{mphase}
}
Here, for a pair of trajectories $\gamma$ and $\gamma^\prime$ from the diagonal
(loop) contribution,  $\msf A \equiv {\int_{\gamma(l)} \mathbf A \cdot d
\mathbf l} / B_z$ is the effective enclosed area, where the integral of the
vector potential~$\mathbf A$ is taken along $\gamma$ (its loop part~$l$), and
$\Phi_0 = hc/e$ is the flux quantum. 

With the Zeeman interaction included, Eqs.~\eq{Mdiag} and~\eq{Mloop} for the
spin modulation factor are no longer valid. In fact, if we distinguish the SO
and Zeeman terms in the effective magnetic field $\mathbf C_\gamma (t) =
\mathbf C_\gamma^{\text{SO}} (t) + \mathbf C_\gamma^Z (t)$, the diagonal (and,
correspondingly, the loop) modulation factor can be written in the form 
\eqn{
  \mc M_{\gamma, \gamma^{-1}} (\mathbf B) = \text{Tr}\, (\hat K_\gamma \hat
  K_{\wtilde \gamma}), 
\label{MZ}
}
where $\wtilde \gamma$ is a fictitious trajectory producing the field $\mathbf
C_{\wtilde \gamma} (t) = \mathbf C_\gamma^{\text{SO}} (t) - \mathbf C_\gamma^Z
(t)$. Clearly, in the absence of SO coupling, we have $\mc M_{\gamma,
\gamma^{-1}} (\mathbf B) = 2s + 1$, i.e., the Zeeman field alone does not
affect the modulation factor. 

In Refs.~\onlinecite{bara93b} and~\onlinecite{rich02} the quantum corrections
to transmission and reflection in the presence of magnetic field were
calculated for an ergodic system. We extend their approach to a system with
SO~interaction. To this end, we consider the \emph{generalized} modulation
factor
\eqn{
  (\mc M \varphi)_{\gamma, \gamma^\prime} \equiv \mc M_{\gamma, \gamma^\prime}
  \, \varphi_{\gamma, \gamma^\prime}.
\label{genmf}
}
The diagonal contribution can be computed using the sum rule~\eq{sumrule}.
First, one averages $(\mc M \varphi)_{\gamma, \gamma^\prime}$ for the
time-reversed pairs of trajectories and loops of a given length~$L$. Thus, the
average is performed over the ensemble of almost \emph{closed} orbits. This
restriction proves very important, since the average modulation factor for
closed and open trajectories is different (see Sec.\ref{rdso}). The average
modulation factor $\wbar{\mc M \varphi}\, (L; \mathbf B)$ is then further
weighted with the length distribution~\eq{PL}. It can be shown that for the
loop contribution in a hyperbolic system with a single Lyapunov exponent holds
effectively the same procedure (see Ref.~\onlinecite{rich02} and
Appendix~\ref{lc}). Hence, the diagonal and the loop \emph{relative} quantum
corrections to reflection and transmission are equal and given~by: 
\aln{
  &\frac {\delta \mc R_{\text{diag}}\, (\mathbf B)} 
  {\delta \mc R_{\text{diag}}^{(0)}} = 
  \frac {\delta \mc R_{\text{loop}}\, (\mathbf B)} 
  {\delta \mc R_{\text{loop}}^{(0)}} =
  \frac {\delta \mc T_{\text{loop}}\, (\mathbf B)} 
  {\delta \mc T_{\text{loop}}^{(0)} } \notag \\
  &= \langle \wbar{\mc M  \varphi}\, (\mathbf B) \rangle_L \notag \\
  &\equiv \frac 1 {L_{\text{esc}}} \int_0^\infty dL\, e^{-L/ L_{\text{esc}}}\,
  \wbar{\mc M  \varphi} \, (L; \mathbf B).
\label{al}
}
The normalization condition~\eq{norm} is preserved due to Eq.~\eq{norm0}.

When the SO and Zeeman interactions are absent, the average modulation
factor~$(2s + 1) \wbar \varphi\, (L; \mathbf B)$ in a chaotic system can be
analytically estimated using the Gaussian  distribution of enclosed
areas~\cite{bara93b}
\eqn{
  P_{\msf A}(\msf A; L) \simeq \frac 1  {\sqrt{2 \pi \msf A_0^2\,  L/ L_b}} \,
  \exp \left( - \frac {\msf A^2} {2 \msf A_0^2\, L/ L_b} \right).
\label{Adist}
}
It depends on a system-specific parameter~$\msf A_0$, a typical area enclosed
by an orbit during one circulation. This distribution does not depend on the
incoming and outgoing channel numbers and is valid for both closed and open
trajectories. The average~\eq{al} yields~\cite{bara93b, rich02}
\eqn{
  \langle \wbar{\mc M  \varphi}\, (\mathbf B) \rangle_L = \frac {2s + 1} {1 +
  \wtilde B^2 L_{\text{esc}} / L_b},
\label{wl}
}
where $\wtilde B = 2 \sqrt 2 \pi B_z \msf A_0 / \Phi_0$. Thus the quantum
corrections have a Lorentzian dependence on the magnetic field. The
result~\eq{wl} is specific to chaotic systems as it depends on
$L_{\text{esc}}$---such a parameter is not relevant to extended disordered
systems, while for regular billiards $P_L(L)$ (introduced earlier) is usually a
power law~\cite{bara93b}. The increase of reflection (decrease of transmission)
for $B_z = 0$ constitutes the effect of \emph{weak localization}. A magnetic
field destroys the time-reversal symmetry and, thereby, the interference
between the mutually time-reversed and loop pairs of paths, thus diminishing
the quantum corrections. 

The SO interaction may turn the constructive interference between the orbit
pairs into destructive one. Since the sign of the quantum corrections in this
case would be reversed, one speaks of \emph{weak antilocalization}. In the
following sections we study the transition from WL to WA and the related
question of spin relaxation.

\section{Spin relaxation} 
\label{sr}

\subsection{General discussion}

Equation~\eq{al} demonstrates that the modulation factor~$\wbar{\mc M \varphi}
\, (L; \mathbf B)$ is a key to calculating the quantum corrections to the
conductance. Hence, we will first examine $\wbar{\mc M \varphi}$ in detail. As
a function of length, this quantity contains information about the average spin
evolution along the trajectories of the system. Here, by the spin evolution
along a trajectory~$\gamma$ we mean the change of the spin propagator~$\hat
K_\gamma (t)$. According to Appendix~\ref{sp}, it can be written in the form
[Eq.~\eq{spprgen}]
\eqn{
  \hat K_\gamma (t) = \exp [- i \hat{\mathbf s} \cdot \bs \eta_\gamma (t)], 
}
and, thus, depends on three real parameters: the rotation angle~$\eta_\gamma
(t)$ and the rotation axis given by the unit vector~$\mathbf m_\gamma (t)
\equiv \bs \eta_\gamma (t) / \eta_\gamma (t)$. Alternatively, one can
parameterize~$\hat K_\gamma$ using the elements of the corresponding SU(2)
matrix [Eq.~\eq{W}]
\eqn{
  W_\gamma (t) = e^{- i \bs \sigma \cdot \bs \eta_\gamma (t)/ 2 } 
  = \left(
  \begin{array}{cc}
    a_\gamma(t) & b_\gamma(t) \\
    - b_\gamma^*(t) & a_\gamma^*(t)
  \end{array}
  \right), 
\label{SU2mat}
} 
which are restricted by the condition $\det W_\gamma = |a_\gamma|^2 +
|b_\gamma|^2 = 1$. The two parameterizations are related by Eq.~\eq{Wrot}.
Clearly, $W_\gamma$ is the matrix representation of $\hat K_\gamma$ for spin $s
= 1/2$. Instead of~$W_\gamma (t)$, we can consider the evolution of the spinor
$\psi_\gamma (t) \equiv \left(a_\gamma (t), - b_\gamma^* (t) \right)^T$,
starting from the spin-up state $\psi_\gamma (0) = (1, 0)^T$. It is
characterized by the spin direction $\mathbf n_\gamma (t) = [\psi_\gamma (t)]^T
\bs \sigma\, \psi_\gamma (t)$ and the overall phase ($\bs \sigma$ is the vector
of Pauli matrices). For spin $s > 1/2$ these are the direction and the phase of
a coherent state (see Sec.~\ref{sost}). Note that $\mathbf n_\gamma (t)$
results from the rotation of $\mathbf n_\gamma (0) = (0, 0, 1)$ by the
angle~$\eta_\gamma (t)$ about~$\mathbf m_\gamma (t)$. Sometimes it is
convenient to represent $W_\gamma (t)$ by a trajectory on the
\emph{three}-dimensional unit sphere~$S^3$. For this purpose we define a
four-dimensional unit vector 
\aln{
  \bs \xi_\gamma (t) &= \bigl( - \text{Im}\, b_\gamma (t), - \text{Re}\,
  b_\gamma(t), - \text{Im}\, a_\gamma(t), \text{Re}\, a_\gamma(t) \bigr)
  \notag \\
  &= \left(\mathbf m_\gamma (t)\, \sin \frac {\eta_\gamma (t)} 2,\, \cos \frac
  {\eta_\gamma (t)} 2 \right). 
\label{xi}
}
The trajectory starts at the ``north pole,'' i.e., $\bs \xi_\gamma (0) = (0, 0,
0, 1)$. 

Using the propagator matrix~\eq{SU2mat}, the modulation factor Eqs.~\eq{Mdiag}
and~\eq{Mloop} for spin~$1/2$ can be expressed~as 
\eqn{
  \mc M = \text{Tr}\, (W^2) = 4\, \xi_4^{\, 2} - 2 = 2 \cos \eta,
\label{M12}
}
where $\xi_4$ is the fourth component of~$\bs \xi$ (in Appendix~\ref{mfas} an
arbitrary~$s$ is considered). To simplify the notation, we dropped the
subscripts labeling the trajectory, and the time~$t$ is the trajectory or loop
time. 

For long orbits one expects that the spin state becomes completely randomized
due to SO interaction, if the particle motion is irregular. This means that all
points $\bs \xi \in S^3$ are equally probable, and, on average, $\wbar
{\xi_4^{\, 2}} = 1/4$ in the limit $L \to \infty$. Hence, for $\mathbf B = 0$
the modulation factor $\wbar{\mc M} (L) \equiv \wbar{\mc M \varphi} \, (L; 0)$
changes with~$L$ from the positive value $\wbar{\mc M} (0) = 2$ to the negative
asymptotic value $\wbar{\mc M} (\infty) = -1$ (cf.\ Ref.~\onlinecite{berg82}).
The stronger the SO interaction, the shorter the length scale~$L_{\mc M}$ of
this change. If $L_{\text{esc}} \ll L_{\mc M}$, i.e., the particle quickly
leaves the cavity, or the SO interaction is weak, then the relative quantum
corrections~\eq{al} are positive, giving rise to~WL. In the opposite case of
strong SO interaction or long dwell times, the relative quantum corrections are
negative, leading to~WA. For an arbitrary spin, $\wbar{\mc M} (L)$ changes from
$\wbar{\mc M} (0) =  2s + 1$ to $\wbar{\mc M} \, (\infty) = (-1)^{2s}$
(Appendix~\ref{mfas}). Thus, WA cannot be observed for an integer spin, at
least, for $L_{\text{esc}} \gg L_{\mc M}$ [$\wbar{\mc M} (L)$ can, in
principle, become negative at intermediate lengths].

For the rest of the paper we will consider the physically most important case
of spin $s = 1/2$.  If the Zeeman interaction is included, then Eq.~\eq{MZ}
yields
\eqn{
  \mc M (\mathbf B) = 4\, \xi_4\, \wtilde \xi_4 - 2\, \bs \xi \cdot \wtilde{\bs
  \xi},
\label{M_Z}
}
where $\wtilde{\bs \xi}$ belongs to the fictitious trajectory~$\wtilde \gamma$.
In the absence of Zeeman coupling, the vectors $\bs \xi$ and~$\wtilde{\bs \xi}$
coincide. Then the negative second term in Eq.~\eq{M_Z} is responsible for the
WA, if the first term is, on average, small enough due to SO~interaction. An
admixture of a moderate Zeeman coupling destroys the correlation between $\bs
\xi$ and~$\wtilde{\bs \xi}$, thereby reducing the average product~$\wbar{\bs
\xi \cdot \wtilde{\bs \xi}}$.  Thus, an external magnetic field suppresses WA
in two ways: the AB~flux breaks down the constructive interference between the
orbital phases and the Zeeman interaction affects the spin modulation factor.
As we know, the former mechanism inhibits the WL, as well. 

The spin propagator $\hat K_\gamma (t)$ can be used not only in the calculation
of the quantum corrections~\eq{al}---it also provides information about the
spin relaxation along classical trajectories, which is of separate interest.
The relaxation of the spin direction can be described by the $z$~component
of the vector~$\mathbf n_\gamma$: 
\eqn{
  n_z = 2\, (\xi_3^{\, 2} + \xi_4^{\, 2}) - 1.
\label{nz}
}
The ensemble average~$\wbar{n_z}\, (L)$ varies from $\wbar{n_z}\, (0) = 1$ to
$\wbar{n_z}\, (\infty) = 0$, if the memory of the initial spin direction is
completely lost for long orbits. The typical length scale of this decay can be
different from~$L_{\mc M}$, because $\wbar{\mc M} (L)$ depends on the
\emph{phase} of the spin state, as well as on its direction. Moreover, the
length scale of~$\wbar{n_z}\, (L)$, as defined by Eq.~\eq{nz}, depends on the
choice of the quantization axis. An \emph{invariant} measure of the spin
relaxation is given by~$\wbar{\xi_4^{\, 2}} (L)$ or, equivalently,
by~$\wbar{\mc M} (L)$. The different relaxation rates of $\wbar{n_z}\, (L)$ and
$\wbar{\mc M} (L)$ are observed in two-dimensional systems with the Rashba and
the Dresselhaus SO coupling (see Secs.~\ref{difsys} and~\ref{rdso}).

\subsection{Example: diffusive systems}
\label{difsys}

In three-dimensional extended diffusive conductors\footnote{Semiclassical
description of diffusive systems implies that the Fermi wavelength is much
smaller than the mean free path.} the directions of the effective magnetic
field~$\mathbf C_\gamma (t)$ before and after a scattering event can be assumed
uncorrelated~\cite{chak86}. We model this by keeping $|\mathbf C| =
\text{const}$ and changing the direction of~$\mathbf C$ randomly at identical
time intervals equal to the elastic scattering time~$\tau$. The spin propagator
for the $j$th time interval is $\hat K_j =  \exp\, (- i \hat{\mathbf s} \cdot
\mathbf m_j\, |\mathbf C|\, \tau)$, $j = 1, 2, \ldots$, where $\mathbf m_j$ is
a random unit vector. The position on~$S^3$ after the first time interval is,
according to Eq.~\eq{xi}, $\bs \xi (\tau) = \left(\mathbf m_1\, \sin \frac
{|\mathbf C|\, \tau} 2,\, \cos \frac {|\mathbf C|\, \tau} 2 \right)$. Thus, a
trajectory on the sphere, starting at the ``north pole,'' traverses an arc of
length~$|\mathbf C|\, \tau /2$ along a randomly chosen great circle. During the
second time interval, the trajectory starts at~$\bs \xi (\tau)$ and moves along
another random great-circle segment, and so on. Clearly, $\bs \xi (t)$~follows
a random walk on~$S^3$. In the continuous limit $|\mathbf C|\, \tau \ll 2 \pi$,
its probability density satisfies a diffusion equation. Solving this equation
(Appendix~\ref{sd}) we find that the average modulation factor for trajectories
of time~$t$,
\eqn{
  \wbar{\mc M}_{\text{diff,\,3D}}\, (t) = 3\, e^{- \frac 1 3 |\mathbf C|^2 \tau
  t} - 1,
\label{Mdiff}
}
and the average spin polarization,
\eqn{
  \wbar{(n_z)}_{\text{diff,\,3D}}\, (t) = e^{- \frac 1 3 |\mathbf C|^2 \tau t},
\label{ndiff}
}
exhibit the same relaxation rate. Note that Eq.~\eq{al} is not valid in
diffusive systems. The modulation factor~\eq{Mdiff} is equivalent to the result
of Eq.~(10.12) of Ref.~\onlinecite{chak86}. 

\begin{figure}[tbp]
  \vspace*{.5cm}
  \includegraphics[width=.95 \linewidth, angle=0]{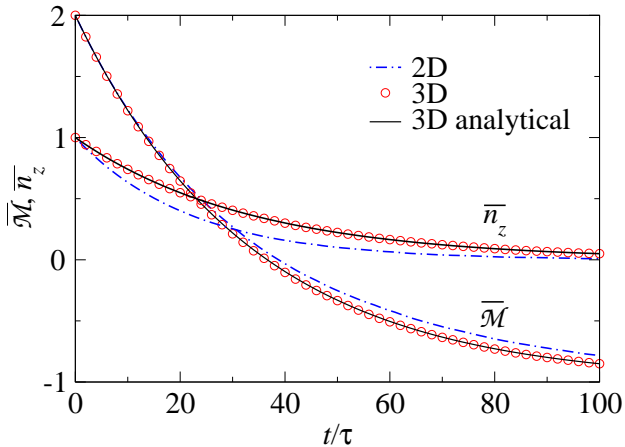}
  \caption{Average spin modulation factor $\wbar{\mc M} (t)$ and spin
  projection $\wbar{n_z}\, (t)$ in diffusive systems. The effective magnetic
  field~$\mathbf C (t)$ changes its direction randomly at equal time
  intervals~$\tau$. Its magnitude is kept constant and is equal to $0.3/\tau$
  in this example. The analytical expressions \eq{Mdiff} and~\eq{ndiff} (solid
  curves) are compared with the results of numerical simulations for $\mathbf
  C$ in two (dash-dotted curves) and three (circles) dimensions. The average
  was performed over $10^5$~random sequences (``trajectories''). 
  \label{modfact_dif_fig}}
\end{figure}

In two-dimensional diffusive systems with Rashba or Dresselhaus interaction it
is reasonable to assume that $\mathbf C$ acquires a random direction in a
two-dimensional plane. In this case the walk on~$S^3$ is not fully random. As
our numerical simulations show (Fig.~\ref{modfact_dif_fig}), the modulation
factor is reasonably well described by Eq.~\eq{Mdiff}. However, the off-plane
polarization~$\wbar{n_z}\, (t)$ relaxes faster in two dimensions [cf.\
Eqs.~(34) and~(35) of Ref.~\onlinecite{piku84}]. This is not surprising, since,
obviously, $\xi_3 (\tau) \equiv 0$ in~2D, but not in~3D.

\section{Rashba and Dresselhaus interaction: Spin relaxation}
\label{rdso}

\subsection{Effective magnetic field}
\label{efmf}

We apply the general theory of the previous sections to ballistic quantum dots
with Rashba~\cite{bych84} and Dresselhaus~\cite{dres55} SO interaction.
Both contributions are usually present in GaAs/AlGaAs heterostructures. Their
strength ratio can be experimentally varied, e.g., by tuning the Rashba SO
strength through an additional gate voltage~\cite{andr97}. When the
two-dimensional electron gas lies in the $(001)$~plane of a zinc-blend lattice,
the effective magnetic field $\hat{\mathbf C} = \hat{\mathbf C}_R +
\hat{\mathbf C}_D$ in the Hamiltonian~\eq{soham} consists~of
\aln{
  &\hat{\mathbf C}_R = \frac {2 \pi} {\Lambda_R}\, \hat{\mathbf v} \times
  \mathbf e_z, 
\label{Rfield}  \\  
  &\hat{\mathbf C}_D = \frac {2 \pi}
  {\Lambda_D}\, (\hat v_x  \mathbf e_x -  \hat v_y \mathbf e_y ),
\label{Dfield}
}
where the $x$ and $y$ axes are chosen along the $[100]$ and $[010]$
crystallographic directions, respectively, and $\hat{\mathbf v} = (\hat{\mathbf
p} - e \mathbf A/c)/M$ is the (Fermi-)velocity operator depending on the
effective mass~$M$. In Eq.~\eq{Rfield} [Eq.~\eq{Dfield}] the Rashba
(Dresselhaus) interaction, usually characterized by the constant
$\alpha_R$~($\alpha_D$), is measured in terms of the inverse spin-precession
length $\Lambda_{R(D)}^{-1} = \alpha_{R(D)} M / \pi \hbar^2$. In billiards the
natural dimensionless parameter is 
\eqn{
  \theta_{R(D)} = 2 \pi L_b/ \Lambda_{R(D)}.
\label{thetaRD}
} 
It signifies the mean spin-precession angle per bounce if only one type of SO
interaction is present.

\begin{figure}[tbp]
  \vspace*{.5cm}
  \includegraphics[width=.5 \linewidth, angle=0]{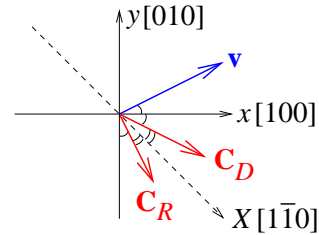}
  \caption{The Rashba effective field~$\mathbf C_R$ is normal to the
  velocity~$\mathbf v$, while the directions of the Dresselhaus field~$\mathbf
  C_D$ and~$\mathbf v$ are symmetric with respect to~$[100]$. Thus, the
  directions of $\mathbf C_R$ and~$\mathbf C_D$ are symmetric with respect
  to~$[1 \bar 1 0]$.
  \label{RDfield_fig}}
\end{figure}

As can be seen from Eq.~\eq{Rfield}, the effective Rashba magnetic
field~$\mathbf C_R (t)$ generated by a particular trajectory points
perpendicular to the velocity~$\mathbf v (t)$. The directions of the
Dresselhaus field~$\mathbf C_D (t)$ and $\mathbf v (t)$ are symmetric with
respect to the $x$~axis [Eq.~\eq{Dfield}]. Hence, $\mathbf C_R (t)$ and
$\mathbf C_D (t)$ always point symmetrically with respect to the $[1 \bar 1
0]$~direction, labeled here by~$X$ (Fig.~\ref{RDfield_fig}). As a consequence,
the total field~$\mathbf C (t)$ is reflected about~$X$ under the exchange
$\Lambda_R \leftrightarrow \Lambda_D$ in Eqs.~\eq{Rfield} and~\eq{Dfield}. This
means that the modulation factor~$\mc M (t)$ and the polarization projections
$n_z (t)$ and $n_X (t)$ are preserved under this transformation. For example,
systems with only Rashba or only Dresselhaus interaction have identical spin
evolution, if the coupling strengths are the same. 

It is sometimes convenient to work in the coordinate frame of $X$ and $Y =
[110]$ (cf.\ Ref.~\onlinecite{alei01}). The projections of the effective
magnetic field on these axes are given by 
\eqn{
  \hat C_X = 2 \pi \hat v_Y / \Lambda_X, \quad 
  \hat C_Y = 2 \pi \hat v_X / \Lambda_Y,
\label{XYfield}
}
where 
\eqn{
  \Lambda_X^{-1} = \Lambda_D^{-1} + \Lambda_R^{-1}, \quad 
  \Lambda_Y^{-1} = \Lambda_D^{-1} - \Lambda_R^{-1}
}
are the effective inverse precession lengths. As above, we can define
dimensionless parameters $\theta_{X(Y)} = 2 \pi L_b/ \Lambda_{X(Y)}$.

\subsection{Numerical study} 
\label{num}

The computation of the spin evolution in billiard cavities is relatively
straightforward, since the classical trajectories there are sequences of
straight segments. If only the \emph{uniform} Rashba and Dresselhaus
interaction is present, the effective magnetic field~$\mathbf C_j$ is constant
along the $j$th~segment (the segment velocity~$\mathbf v_j$ is constant,
moreover, its magnitude~$v$ is the same for all~$j$ due to the energy
conservation). The spin-propagator matrix~\eq{SU2mat} for a trajectory, 
\eqn{
  W \equiv e^{- i \bs \sigma \cdot \bs \eta / 2} = W_l \cdots W_1, 
\label{Wtr}
}
is a product of the respective matrices
\eqn{
  W_j \equiv e^{- i \bs \sigma \cdot \bs \eta_j / 2} = e^{- i \bs \sigma \cdot 
  \mathbf C_j t_j/2} \quad (j = 1,\, \ldots, l)
}
for the $l$ orbit segments. In practice, it is convenient to remove the
velocity dependence in Eqs.~\eq{XYfield} by using the displacement~$\Delta
\mathbf r_j \equiv \Delta X_j\, \mathbf e_X + \Delta Y_j\, \mathbf e_Y =
\mathbf v_j\, t_j$ instead of the segment time~$t_j$. Thereby the rotation
vector can be expressed as
\eqn{
  \bs \eta_j = 2 \pi \left( \frac {\Delta Y_j} {\Lambda_X}\, \mathbf e_X + 
  \frac {\Delta X_j} {\Lambda_Y}\, \mathbf e_Y \right).
\label{etaj}
}
It follows from this equation that rescaling of the system size and the
spin-precession lengths by the same factor does not change the spin relaxation.
In other words, given the shape of the billiard, the averages $\wbar{n_z}$
and~$\wbar{\mc M}$ as functions of $L/L_b$ (computed below) depend only on the
angles $\theta_R$ and~$\theta_D$. 

\begin{figure}[tbp]
  \vspace*{.5cm}
  \includegraphics[width=.95 \linewidth, angle=0]{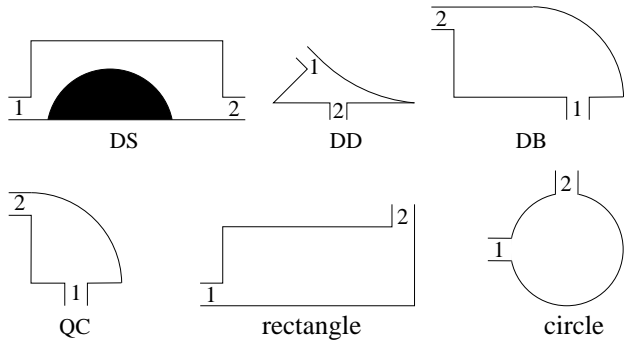}
  \caption{Billiard geometries: desymmetrized Sinai~(DS) billiard,
  desymmetrized diamond~(DD) billiard, desymmetrized Bunimovich~(DB) stadium
  billiard, quarter circle~(QC), rectangle, and circle. The leads are numbered
  for future reference.
  \label{bill_fig}}
\end{figure}

We performed a systematic numerical study of spin relaxation for several
billiard geometries (Fig.~\ref{bill_fig}) representative for systems with
chaotic and integrable classical dynamics. The desymmetrized Sinai~(DS)
billiard, the desymmetrized diamond~\cite{muel03}~(DD) billiard, and the
desymmetrized Bunimovich~(DB) stadium billiard represent chaotic cavities. The
quarter circle~(QC), rectangle, and circle are integrable. The average spin
relaxation is computed for the \emph{closed} versions of these billiards.

\begin{figure}[tbp]
  \vspace*{.5cm}
  \includegraphics[width=.95 \linewidth, angle=0]{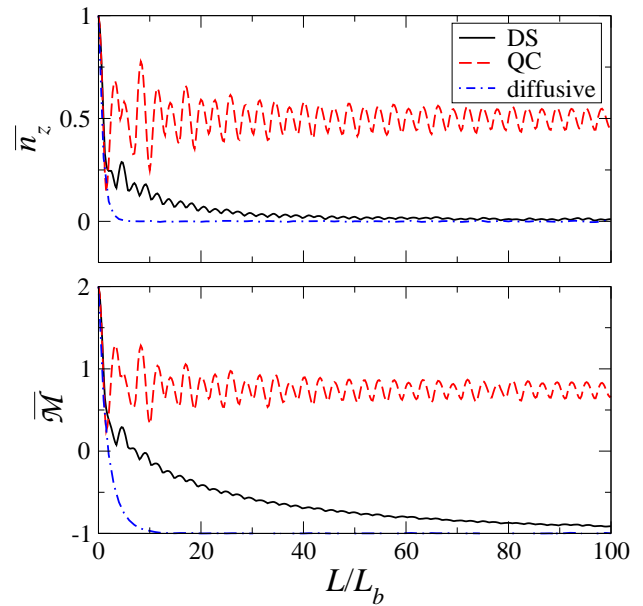}
  \caption{Average spin projection~$\wbar{n_z} (L)$ and modulation
  factor~$\wbar{\mc M} (L)$ for the closed chaotic~DS (solid curves) and
  integrable~QC (dashed curves) billiard (see Fig.~\ref{bill_fig}). Each data
  point represents the average over 50,000 open trajectories with random
  initial values at the boundary. The numerical results for a two-dimensional
  extended diffusive system (dash-dotted curves), where $L_b$ is identified
  with the elastic mean free path~$v \tau$, are shown for comparison. The
  SO-coupling strength is $\theta_R/ 2 \pi = 0.2$ and $\theta_D = 0$.
  \label{cha_int_open_fig}}
\end{figure}

\begin{figure}[tbp]
  \vspace*{.5cm}
  \includegraphics[width=.95 \linewidth, angle=0]{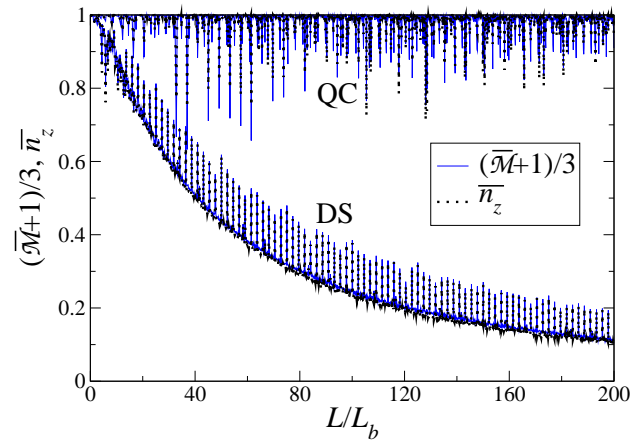}
  \caption{Average spin projection~$\wbar{n_z} (L)$ (dotted curves) and 
  rescaled modulation factor~$[\wbar{\mc M} (L) + 1]/3$ (solid curves) for the
  closed chaotic~DS and integrable~QC billiard. Each data point represents the
  average over 5,000 closed trajectories started at random at the
  boundary. The SO-coupling strength is $\theta_R/ 2 \pi = 0.2$ and
  $\theta_D = 0$.
  \label{cha_int_closed_fig}}
\end{figure}

Figures \ref{cha_int_open_fig} and~\ref{cha_int_closed_fig} depict the average
spin relaxation described by $\wbar{n_z} (L)$ and its invariant counterpart,
$\wbar{\mc M} (L)$, for the  chaotic~DS and integrable~QC billiard. The average
is performed over ensembles of open (Fig.~\ref{cha_int_open_fig}) and closed
(Fig.~\ref{cha_int_closed_fig}) trajectories of length~$L$ starting at random
position at the boundary with a random boundary component of the velocity. The
strength of the Rashba interaction is chosen as $\theta_R/ 2 \pi = 0.2$, and
the Dresselhaus interaction is absent (or, equivalently,  $\theta_D/ 2 \pi =
0.2$ and $\theta_R = 0$). 

In Fig.~\ref{cha_int_open_fig}, the numerical results for an extended
two-dimensional diffusive system, where $L_b$ is identified with the elastic
mean free path~$v \tau$, are shown for comparison. We observe that on the scale
of $L \sim L_b$ the spin relaxation is the same in all three examples. Indeed,
before the first collision with the boundary or a scatterer, the particle moves
along a straight line, irrespective of the system it belongs~to. On longer
length scales relaxation in an extended diffusive system is much stronger than
in confined systems. Moreover, in the integrable billiard saturation takes
place.\footnote{It was argued in Ref.~\onlinecite{chan04b} that the saturation
is due to the nearly periodic orbits, which have a finite measure in integrable
systems.} We also note that $\wbar{n_z} (L)$ in the chaotic billiard, similarly
to the diffusive system (Sec.~\ref{difsys}), relaxes to its asymptotic value
faster than~$\wbar{\mc M} (L)$. 

The ensemble of closed orbits (Fig.~\ref{cha_int_closed_fig}) is responsible
for the quantum corrections to transmission and reflection
(Sec.~\ref{semland}). We find that the relaxation in this case is much slower
than for the ensemble of arbitrary trajectories. Remarkably, the spin
projection $\wbar{n_z}$ and the rescaled modulation factor $(\wbar{\mc M} +
1)/3$ are hardly distinguishable.  

\begin{figure}[tbp]
  \vspace*{.5cm}
  \includegraphics[width=.95 \linewidth, angle=0]{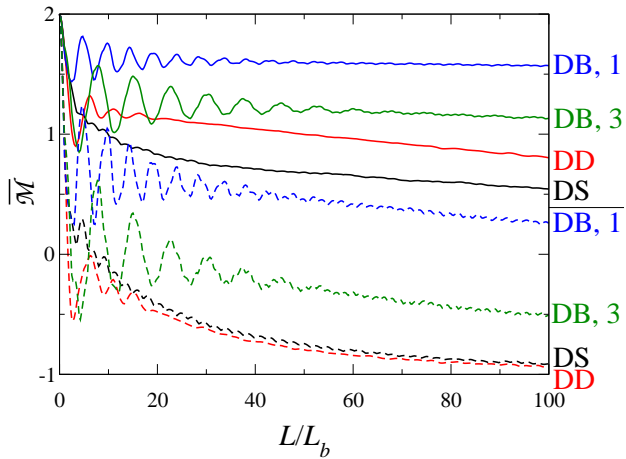}
  \caption{The average modulation factor~$\wbar{\mc M} (L)$ for the closed
  chaotic DS, DD, and DB billiards. In the latter case the ratio of the upper
  straight side to the radius was taken to be~1 (DB,~1) and~3 (DB,~3). Each
  data point represents the average over 50,000 open trajectories started at
  random at the boundary. The SO-coupling strength is $\theta_R/ 2 \pi = 0.1$
  (four upper curves) and $0.2$ (four lower curves), while $\theta_D = 0$. 
  \label{cha_open_fig}}
\end{figure}

\begin{figure}[tbp]
  \vspace*{.5cm}
  \includegraphics[width=.95 \linewidth, angle=0]{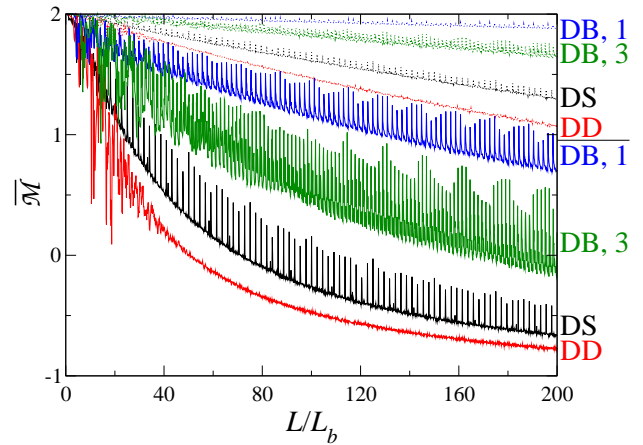}
  \caption{Same as in Fig.~\ref{cha_open_fig}, but for the ensemble of closed
  orbits started at random at the boundary. Each data point represents the
  average over 5,000 trajectories.
  \label{cha_closed_fig}}
\end{figure}

The spin evolution in several chaotic systems is compared in
Figs.~\ref{cha_open_fig} and~\ref{cha_closed_fig} for the ensembles of open and
closed trajectories, respectively. All the billiards show a qualitatively
similar behavior. The DS and DD billiards have about the same relaxation rate.
In the DB~billiards the relaxation rate grows continuously, starting from zero,
as the ratio of the upper straight side to the radius increases. 

\begin{figure}[tbp]
  \vspace*{.65cm}
  \includegraphics[width=.95 \linewidth, angle=0]{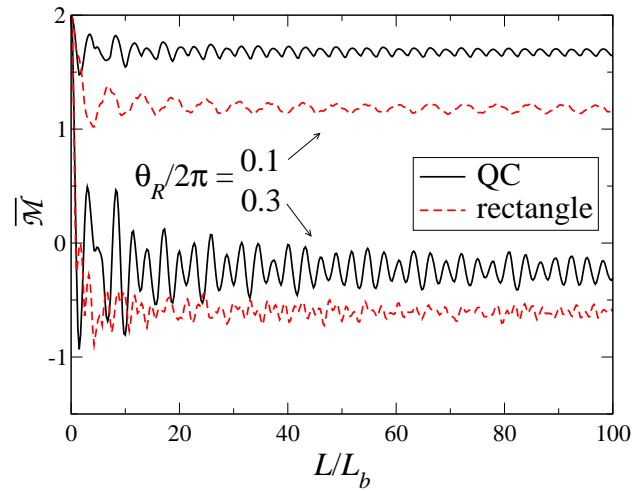}
  \caption{The average modulation factor~$\wbar{\mc M} (L)$ for the closed
  integrable QC (solid curves) and rectangle (dashed curves) billiards. Each
  data point represents the average over 50,000 open trajectories started at
  random at the boundary. The SO-coupling strength is $\theta_R/ 2 \pi = 0.1$
  and~$0.3$, while $\theta_D = 0$. 
  \label{int_open_fig}}
\end{figure}

In Fig.~\ref{int_open_fig} the modulation factor averaged over the open
trajectories is presented for the integrable QC and rectangle billiard. Both
systems are characterized by the saturation of spin relaxation and persistent
long-time oscillations. The saturation level decreases down to $-1$ as the
SO coupling becomes stronger. The circle billiard, which shows a
non-typical relaxation pattern, is considered in Sec.~\ref{lssm}. 

\begin{figure}[tbp]
  \vspace*{.5cm}
  \includegraphics[width=.95 \linewidth, angle=0]{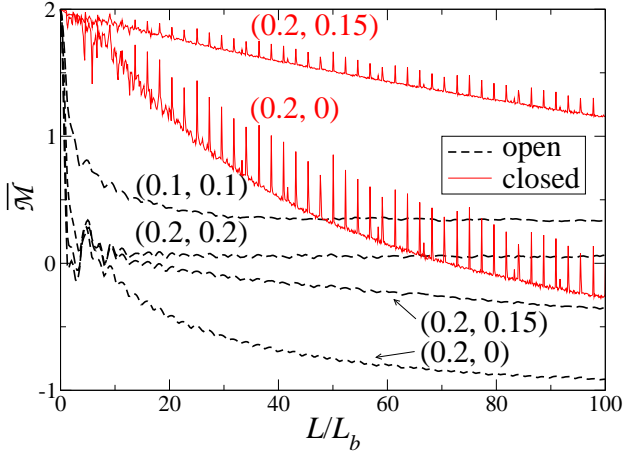}
  \caption{The average modulation factor~$\wbar{\mc M} (L)$ for the closed
  DS~billiard at different strengths of the Rashba and Dresselhaus interaction.
  Each data point represents the average over 50,000 open (dashed curves) or
  5,000 closed (solid curves) trajectories started at random at the boundary.
  The values of $(\theta_R/ 2 \pi, \theta_D/ 2 \pi)$ are shown on the graph.
  The leads are parallel to the $[100]$ direction. Note that $\mc M = 2$ for
  closed orbits, when $\theta_R = \theta_D$.
  \label{RD_fig}}
\end{figure}

When the Rashba and the Dresselhaus couplings work simultaneously, they
mutually counteract their effects on the spin relaxation\footnote{The interplay
between the Rashba and the Dresselhaus terms in extended diffusive systems was
studied in Ref.~\onlinecite{aver99} (spin relaxation) and in
Ref.~\onlinecite{piku95} (weak antilocalization).} (Fig.~\ref{RD_fig}). In the
extreme case $\Lambda_R = \Lambda_D$, i.e., $\Lambda_Y^{-1} = 0$, the effective
field~$\mathbf C$ is always parallel to the $X$~axis\footnote{In this this case
the system possesses an additional conserved quantity~\cite{schl03}.}
(Fig.~\ref{RDfield_fig}). Hence, the propagator matrices in Eq.~\eq{Wtr}
commute, and the rotation vector becomes
\eqn{
  \bs \eta = \sum_{j=1}^l \bs \eta_j = 2 \pi\, (\Delta Y / \Lambda_X)\, \mathbf
  e_X,
}
where $\Delta Y = \sum_{j=1}^l \Delta Y_j$ is the $Y$~displacement for the
trajectory. According to Eq.~\eq{M12}, the modulation factor is then
\eqn{
  \mc M = 2 \cos\, (2 \pi\, |\Delta Y| / \Lambda_X).
}
On the long-length scale $L \gg L_b$, $|\Delta Y|$ varies from orbit to orbit
between $0$ and the system size. Clearly, the average~$\wbar{\mc M}$ should be
independent of the orbit length~$L$. This explains the saturation in
Fig.~\ref{RD_fig}. The saturation level decreases from $2$ to $0$ as $\theta_R$
changes from $0$ to~$\infty$. For closed orbits we have $\Delta Y = 0$, and,
therefore, $\mc M = 2$. If the spin is initially polarized in the
$X$~direction, the polarization does not change with~$L$, i.e., $n_X = 1$. On
the other hand, Eqs.~\eq{xi} and~\eq{nz} yield $n_z = \mc M / 2$ if the
trajectory starts with $n_z = 1$. This demonstrates that the spin-relaxation
measure $\wbar{n_z} (L)$ depends on the choice of the quantization axis, as was
mentioned in Sec.~\ref{sr}.

\subsection{Limit of slow spin motion}
\label{lssm}

Above we presented a number of numerical observations regarding the average
spin relaxation in two-dimensional billiards.  A further insight into the
connection of the spin dynamics to the characteristics of the orbital motion
can be gained within the limit of slow spin. In this approximation the period
of spin precession is, by definition, much longer than the time scale on which
the orbital momentum changes. In billiards this requirement takes the form
$|\mathbf C| L_b / v \ll 2 \pi$, or,  $\theta_{X,Y} \ll 2 \pi$.

\subsubsection{Rotation-angle expansion}

The rotation angle~$\bs \eta$ [Eq.~\eq{Wtr}] contains all the information about
the spin evolution along a particular trajectory. It essentially depends on the
rotation angles~$\bs \eta_j$ for the straight segments of the trajectory. In
turn, the angles~$\bs \eta_j$ are directly related to the orbital displacements
via Eq.~\eq{etaj}. Thus, a more explicit expression of $\bs \eta$ in terms of
the $\bs \eta_j$ is desirable in order to establish the link between the
geometry of orbital motion and the spin rotation. For this purpose we employ
the Baker-Campbell-Hausdorff (BCH) formula~\cite{vara74, wilc67} for the
product of the exponentials of two matrices (or operators),
\eqn{
  \exp\, (P) \exp\, (Q) = \exp \left( \sum_{i=1}^\infty R_i \right),
\label{BCH}
}
where $R_i$'s are homogeneous polynomials of degree~$i$ in $P$ and~$Q$. The
first three of them are given by
\aln{
  &R_1 = P + Q, \\
  &R_2 = \frac 1 2 [P, Q], \\
  &R_3 = \frac 1 {12} [P - Q, [P, Q]]
}
($[\, ,]$ denotes a commutator). The BCH formula can be used to calculate the
product of the segment propagators $W_j$ in Eq.~\eq{Wtr}. However, only in the
limit of slow spin, when $|\bs \eta_j| \ll 1$, the first few
contributions~$R_i$ are sufficient. The rotation angle for a trajectory, 
\eqn{
  \bs \eta = \bs \eta^{(0)} + \delta \bs \eta^\perp + \delta \bs
  \eta^\parallel,
}
comprises three parts. The lowest-order term,
\eqn{
  \bs \eta^{(0)} = \sum_{j=1}^l \bs \eta_j = 2 \pi \left( \frac {\Delta Y}
  {\Lambda_X}\, \mathbf e_X +  \frac {\Delta X} {\Lambda_Y}\, \mathbf e_Y
  \right),
\label{eta0}
}
is a vector sum of the segment rotation angles ($\Delta X = \sum_{j=1}^l \Delta
X_j$). The correction normal to the billiard plane, 
\eqn{
  \delta \bs \eta^\perp =  \frac {(2 \pi)^2} {\Lambda_X \Lambda_Y}\, \msf A \,
  \mathbf e_z +  \mc O (\Lambda_{X, Y}^{\, -4}),
\label{etaperp}
}
is proportional to the effective enclosed area~$\msf A$ (Appendix~\ref{ac}).
For closed orbits, $\msf A$~coincides with the area defined below
Eq.~\eq{mphase} for a returning orbit or a loop. In an open trajectory, $\msf
A$~is the area of its ``closure'' obtained by connecting the endpoints with a
straight line. The in-plane correction~$\delta \bs \eta^\parallel$ is of the
order~$\Lambda_{X, Y}^{\, -3}$. In general, the in-plane (normal) contribution
to~$\bs \eta$ contains odd (even) powers of~$\bs \eta_j$. 

Expressions \eq{eta0} and~\eq{etaperp} help to interpret some of the results of
Sec.~\ref{num}:
\begin{trivlist}
\item (i) The suppression of spin relaxation along open trajectories in
confined systems, as compared to extended diffusive systems
(Fig.~\ref{cha_int_open_fig}), can be deduced from the expansions 
\eqn{
  \left.
  \begin{array}{c}
    \mc M / 2 \\
    n_z
  \end{array}
  \right\} = 1 - \frac 1 2 \left|\bs \eta^{(0)}\right|^2 +  \mc O (\Lambda_{X,
  Y}^{\, -4}),
\label{Mnzopen}
}
that follow from Eqs.~\eq{M12} and~\eq{nz}. In billiards $\wbar{\left|\bs
\eta^{(0)}\right|^2}$~increases from zero to its saturation value of the order
of~$(\theta_{X,Y})^{\, 2}$ on the length scale of the system size. The further
relaxation on the scale $L \gtrsim L_b$ is due to the $\Lambda_{X, Y}^{\,
-4}$-order terms. In diffusive systems, on the other hand, $\wbar{\left|\bs
\eta^{(0)}\right|^2}$ grows linearly with length. 

\item (ii) For the closed orbits one sets $\bs \eta^{(0)} = 0$. Hence, the
short-scale relaxation does not show up in the ensemble averages $\wbar{\mc M}
(L)$ and~$\wbar{n_z} (L)$ (Fig.~\ref{cha_int_closed_fig}). If the in-plane
contribution~$\delta \bs \eta^\parallel$ is neglected, i.e., $|\delta \bs
\eta^\parallel| \ll 1$, the modulation factor becomes 
\eqn{
  \mc M \simeq 2 \cos |\delta \bs \eta^\perp|,
}
and the generalized modulation factor is
\eqn{
  \mc M \varphi \simeq \sum_\pm \exp \left[ i \msf A \left( \frac {4 \pi B_z}
  {\Phi_0} \pm \frac {(2 \pi)^2} {\Lambda_X \Lambda_Y} \right) \right].
\label{Mweak}
}
In a chaotic system, averaging with the area distribution~\eq{Adist} yields
\eqn{
  \wbar{\mc M \varphi}\, (L; B_z) \simeq \sum_\pm \exp \left[ - ( \wtilde B \pm
  \wtilde \Lambda^{-2})^2 \frac L {L_b} \right],
\label{Mphich}
}
where $\wtilde \Lambda^{-2} \equiv 2 \sqrt 2 \pi^2 \msf A_0 / \Lambda_X
\Lambda_Y$. Clearly, the normal contribution~$\delta \bs \eta^\perp$ alone
cannot make $\wbar{\mc M \varphi}$ negative. For sufficiently large
lengths~$L$, the components $\delta \bs \eta^\perp$ and~$\delta \bs
\eta^\parallel$ become comparable and reverse the sign of the modulation
factor. With the help of Eqs.~\eq{xi} and~\eq{nz}, we obtain the spin
polarization 
\aln{
  n_z &= 1 - 2\, (m_x^{\, 2} + m_y^{\, 2}) \left(\sin \frac \eta 2 \right)^2 
  \notag \\
  &\simeq 1 - 2 \frac {|\delta \bs \eta^\parallel|^2} {|\delta \bs
  \eta^\perp|^2} \left(\sin \frac {|\delta \bs \eta^\perp|} 2 \right)^2.
}
The $\delta \bs \eta^\parallel$~component is responsible for the rotation of an
initially spin-up state. This means that for sufficiently weak SO
interaction and short~$L$, $\wbar{n_z}$ relaxes slower than~$\wbar{\mc M }$
(Fig.~\ref{mfpol_weak_fig}): the relaxation rates are of the order of
$\Lambda_{X, Y}^{\, -6}$ and~$\Lambda_{X, Y}^{\, -4}$, respectively. For
stronger interaction the difference is not noticeable
(Fig.~\ref{cha_int_closed_fig}).

\begin{figure}[tbp]
  \vspace*{.5cm}
  \includegraphics[width=.95 \linewidth, angle=0]{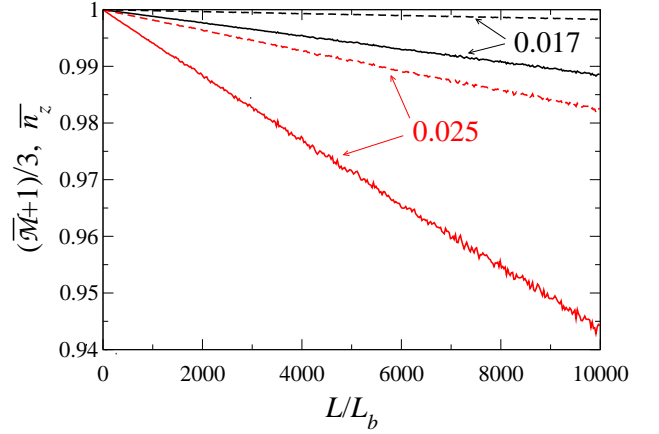}
  \caption{Average spin projection~$\wbar{n_z} (L)$ (dashed curves) and 
  rescaled modulation factor~$[\wbar{\mc M} (L) + 1]/3$ (solid curves) for the
  closed chaotic~DS billiard. Each data point represents the average over 
  15,000 closed trajectories started at random at the boundary. The SO-coupling
  strength $\theta_R/ 2 \pi$ is shown in the graph and $\theta_D = 0$.
  \label{mfpol_weak_fig}}
\end{figure}

\item (iii) In integrable billiards the strong area cancellation~\cite{bara93b}
along trajectories contributes to the saturation of spin relaxation. The circle
billiard makes an exception: here all the trajectories accumulate area linearly
in time. For an orbit having the shortest distance~$r$ from the center, the
enclosed area is, on average, $\msf A \simeq \pm rL/2$ (the signs denote the
two senses of rotation). Averaging the modulation factor~\eq{Mweak} over
all~$\pm r$, i.e., assuming that all angular momenta are equally possible, we
find~\cite{zait04}
\aln{
  &\wbar{\mc M \varphi}\, (L; B_z) \simeq \sum_\pm \frac {\sin \alpha_\pm}
  {\alpha_\pm}, \notag \\ 
  &\alpha_\pm \equiv 2 \pi R L \left( \frac {B_z} {\Phi_0} \pm \frac \pi
  {\Lambda_X \Lambda_Y} \right),
\label{Mcirc}
}
where $R$ is the radius of the circle. [Note that a similar average yields $L_b
= \pi R/2$, in accordance with Eq.~\eq{Lb}.] Surprisingly, for a sufficiently
small SO coupling, the average over open trajectories agrees better
with Eq.~\eq{Mcirc}, than the average over the closed orbits started at the
boundary [Fig.~\ref{circle_fig}~(a)]. This happens because the $\bs
\eta^{(0)}$~contribution is small, but different angular momenta are not
equally represented in the ensemble of closed orbits. As well as
Eq.~\eq{Mphich}, Eq.~\eq{Mcirc} does not describe the full relaxation at large
SO~interaction [Fig.~\ref{circle_fig}~(b)].

\begin{figure}[tbp]
  \vspace*{.5cm}
  \includegraphics[width=.95 \linewidth, angle=0]{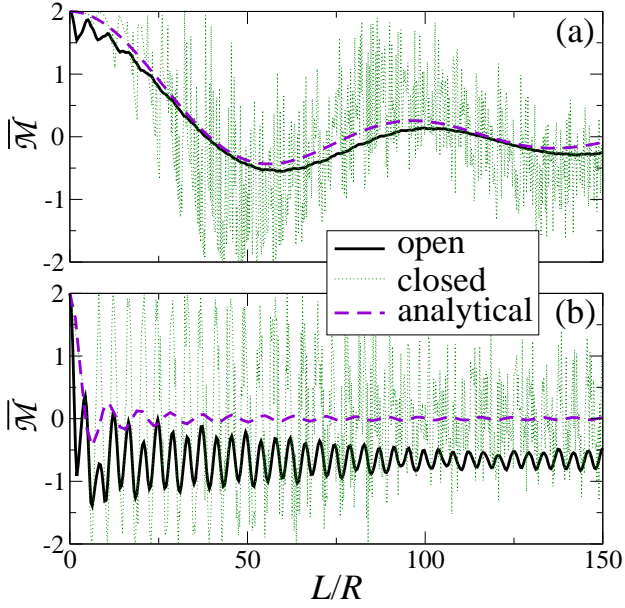}
  \caption{The average modulation factor~$\wbar{\mc M} (L)$ for the closed
  circle billiard of radius~$R$. Each data point represents the average over
  50,000 open (solid curves) or 5,000 closed (dotted curves) trajectories
  started at random at the boundary. The results are compared with the
  analytical expression~\eq{Mcirc} (dashed curves). The SO-coupling strength is
  $\theta_R/ 2 \pi = \pi R /2 \Lambda_R= 0.1$~(a) and~$0.3$~(b), while
  $\theta_D = 0$.
  \label{circle_fig}}
\end{figure}
\end{trivlist}

\subsubsection{Unitary transformation of the Hamiltonian}

It is possible to transform the Rashba-Dresselhaus Hamiltonian [Eqs.~\eq{soham}
and~\eq{XYfield}] to the form in which the SO interaction is weaker by a factor
of~$\theta_{X, Y} \ll 1$ (Refs.~\onlinecite{alei01} and~\onlinecite{crem03}).
Considering the case $\mathbf B = 0$, we start with the Hamiltonian
\eqn{
  \hat H = \frac {\hat{\mathbf p}^2} {2M} + \frac {\pi \hbar} M \left( \frac
  {\hat p_Y\, \sigma_X} {\Lambda_X} + \frac {\hat p_X\, \sigma_Y} {\Lambda_Y} 
  \right)
}
and apply the unitary transformation
\eqn{
  \hat V = \exp \left[ -i \pi \left( \frac {Y \sigma_X} {\Lambda_X} + \frac {X
  \sigma_Y} {\Lambda_Y} \right) \right]
}
to it. In the limit of slow spin motion, the exponent can be expanded in 
powers of~$\Lambda_{X, Y}^{\, -1}$. Keeping only the terms quadratic
in~$\theta_{X, Y}$ and linear in~$\hbar / \Lambda_{X, Y}$ [due to the
weak-coupling assumption~\eq{wcl}], we obtain the new Hamiltonian 
\mul{
  \hat{\wtilde H} = \hat V^\dag \hat H \hat V 
  = \frac {\hat{\mathbf p}^2} {2M}  - \left( \mathbf A^{(1)} + \mathbf
  A^{(2)} \right) \cdot \frac {\hat{\mathbf p}} M \\ + \mc O \left[\theta_{X,
  Y}^{\, 3}, \left(\frac \hbar  {\Lambda_{X, Y}} \right)^2 \right].
\label{Htrans}
}
It contains the spin-dependent ``vector potentials''
\aln{
  &\mathbf A^{(1)} = - \frac{\pi^2 \hbar} {\Lambda_X \Lambda_Y}\,
  (\mathbf e_z \times \mathbf r)\, \sigma_z,  
  \label{A1} \\
  &\mathbf A^{(2)} =  \frac{4 \pi^3 \hbar} {3\, \Lambda_X \Lambda_Y}\, (\mathbf
  e_z \times \mathbf r) \left( \frac {X \sigma_X} {\Lambda_Y} - \frac {Y
  \sigma_Y} {\Lambda_X} \right)
}
of the order of $(\hbar / \Lambda_{X, Y})\, \theta_{X, Y}$ and~$(\hbar /
\Lambda_{X, Y})\, \theta_{X, Y}^{\, 2}$, respectively.

As any two Hamiltonians connected by a unitary transformation, $\hat H$~and
$\hat{\wtilde H}$ must yield the same total transmission~$\mc T$ and
reflection~$\mc R$. While $\hat H$ and $\hat{\wtilde H}$ have identical orbital
parts, the SO interaction in the latter is much weaker. It is
important, therefore, to understand the role of the transformation~$\hat V$
within the framework of the semiclassical Landauer formula developed in the
previous sections. Working in the weak-coupling regime~\eq{wcl} and treating
the orbital motion semiclassically, we consider the unitary transformation 
\eqn{
  V (t) = \exp \left\{ -i \pi \left[ \frac {Y (t)\, \sigma_X} {\Lambda_X} +
  \frac {X (t)\, \sigma_Y} {\Lambda_Y} \right] \right\}
}
acting in the spin space. It is generated by a classical trajectory $[\mathbf r
(t), \mathbf v (t)]$. The spin-propagator matrix~\eq{SU2mat} transforms as 
\eqn{
  \wtilde W (t) = V^\dag (t)\, W (t)\, V (0).
\label{Wtrans}
} 
[This property follows from the spinor transformation $\wtilde \psi (t) =
V^\dag (t)\, \psi (t)$ and the definition $\wtilde \psi (t) = \wtilde W (t)\,
\wtilde \psi (0)$.] For a closed orbit of time~$t$, we have $V (t) = V (0)$.
Hence, the modulation factor for a pair of trajectories from the diagonal or
the loop contribution (Sec.~\ref{semland}),
\eqn{
  \wtilde {\mc M} (t) \equiv \text{Tr}\, \left[\wtilde W (t)\right]^2 = \mc M
  (t),
\label{Mtilde}
}
is preserved under the transformation. Thus, $\hat H$~and $\hat{\wtilde H}$
return the same conductance also in the semiclassical
approximation.\footnote{The partial coefficients $|t_{nm}|^2$ and~$|r_{nm}|^2$
are invariant, as well.} This result demonstrates the significance of closed
trajectories for the transmission and reflection. 

According to Eq.~\eq{Mtilde}, the modulation factor for closed orbits can be
determined directly from the semiclassical version of~$\hat{\wtilde H}$.
Neglecting~$\mathbf A^{(2)}$, we compute the propagator~as 
\aln{
  \wtilde W (t) &\simeq \exp \left( \frac i \hbar \int_0^t \mathbf A^{(1)}
  \cdot \mathbf  v\, dt^\prime \right) \notag \\
  &= \exp \left( -i\, \frac {2 \pi^2} {\Lambda_X \Lambda_Y}\, \msf A\, \sigma_z
  \right).
}
Then Eq.~\eq{etaperp} for the rotation angle and Eq.~\eq{Mweak} for the
modulation factor (with $\mathbf B = 0$) follow. 

For the open trajectories, the leading-order (in $\Lambda_{X, Y}^{\, -1}$)
contribution from Eq.~\eq{Wtrans} is
\aln{
  &W (t) \simeq V (t) V^\dag (0) \notag \\
  &\simeq \exp \left\{ -i \pi \left[ \frac {Y(t) - Y(0) } {\Lambda_X}\,
  \sigma_X + \frac {X(t) - X(0)} {\Lambda_Y}\, \sigma_Y \right] \right\}. 
}
As a consequence, Eqs.~\eq{eta0} and~\eq{Mnzopen} are obtained.

\section{Rashba and Dresselhaus interaction: Quantum corrections to 
transmission and reflection}
\label{rdqc}

\subsection{Results of the semiclassical theory}

As was shown in the previous sections, the average relaxation of the spin
modulation factor as a function of orbit length is closely related to the type
of the classical dynamics of the system. In view of Eq.~\eq{al}, the effect of
the classical orbital motion is also felt in the magnitude and \emph{sign} of
the quantum correction to the conductance. 

In the examples below the following numerical procedure is employed to
calculate the relative quantum corrections: The trajectories are started at the
boundary within a given lead. Their initial coordinate and velocity component
at the lead cross section is chosen at random. If an orbit returns to the same
lead, the generalized modulation factor~\eq{genmf} for this orbit and its
time-reversed partner is recorded. The average over these modulation factors
yields the diagonal contribution\footnote{The requirement that the incoming and
outgoing channels be the same was neglected.}~$\delta \mc R_{\text{diag}}\,
(\mathbf B) / \delta \mc R_{\text{diag}}^{(0)}$, which, by virtue of
Eq.~\eq{al}, is equal to the total relative quantum corrections $\delta \mc
R(\mathbf B)/ \delta \mc R^{(0)} = \delta \mc T(\mathbf B)/ \delta \mc T^{(0)}$
for a chaotic system. For non-chaotic billiards there exist no analytical
approach to the loop statistics. Hence, there we use the diagonal part as an
estimate for  $\delta \mc R(\mathbf B)/ \delta \mc R^{(0)}$. 

\begin{figure}[tbp]
  \vspace*{.5cm}
  \includegraphics[width=.95 \linewidth, angle=0]{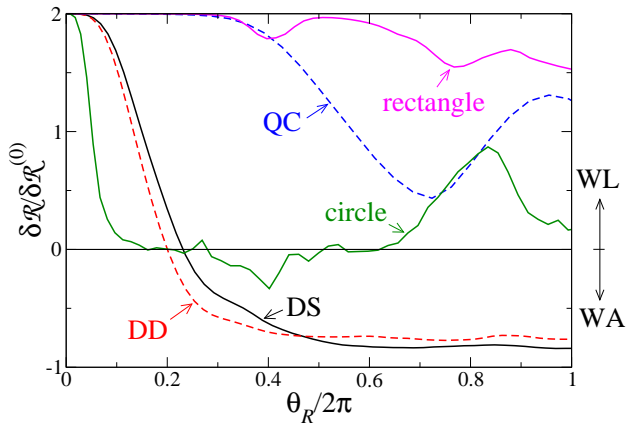}
  \caption{Relative quantum correction to reflection (or transmission)
  coefficients \emph{vs.}\ Rashba-interaction strength~$\theta_R$ ($\theta_D =
  0$). The data are shown for the chaotic (DS and DD) and integrable (QC,
  rectangle, and circle) billiards with $\msf P_c /(w + w^\prime) = 90$. Each
  data point represents the average over 50,000 orbits. The details on the
  numerical procedure are given in the text. 
  \label{antiloc_cha_int_fig}}
\end{figure}

In Fig.~\ref{antiloc_cha_int_fig} the relative quantum correction is shown as a
function of the Rashba-interaction strength~$\theta_R$ (while $\theta_D = 0$)
for several billiards. Here and in the following figures the results for
lead~``1'' (Fig.~\ref{bill_fig}) are presented. All the billiards have the same
ratio $\msf P_c /(w + w^\prime)$, which in chaotic systems gives the escape
length in units of~$L_b$. As $\theta_R$ increases, the relative quantum
corrections in chaotic cavities decrease and, eventually, change sign.  The
level $\delta \mc R/ \delta \mc R^{(0)} = 0$ corresponds to the WL-WA
transition. In the integrable QC and rectangle billiards no WA occurs, while
the circle billiard shows a non-typical pattern.  

\begin{figure}[tbp]
  \vspace*{.5cm}
  \includegraphics[width=.95 \linewidth, angle=0]{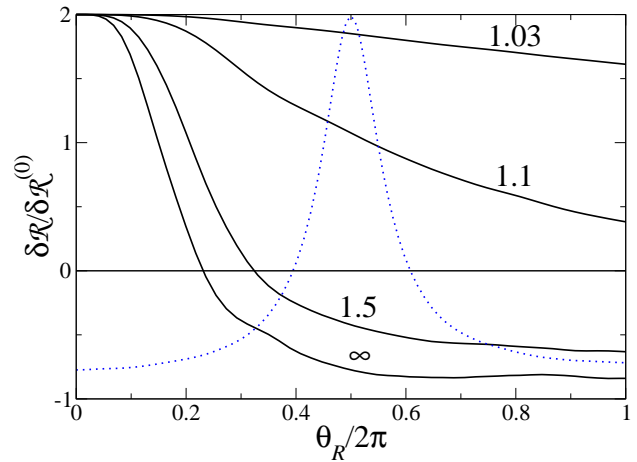}
  \caption{The relative quantum correction to reflection (or transmission)
  coefficients as a function of~$\theta_R$ for the chaotic DS billiard with
  $\msf P_c /(w + w^\prime) = 90$. Solid curves: fixed ratios $\theta_R/
  \theta_D$ (shown in the graph). Dotted curve: fixed $\theta_D = \pi$. Each
  data point represents the average over 50,000 orbits. 
  \label{antiloc_RD_fig}}
\end{figure}

The interplay between Rashba and Dresselhaus interaction in a chaotic billiard
is investigated in Fig.~\ref{antiloc_RD_fig}. As was already noted in
Sec.~\ref{num}, the two interactions tend to compensate each other (thus
suppressing the~WA) as their ratio gets closer to unity. When $\theta_R =
\theta_D$, there is no manifestation of SO interaction in the conductance
[$\delta \mc R(\mathbf B = 0)/ \delta \mc R^{(0)} \simeq 2$], since,
effectively, no spin relaxation is associated with closed trajectories. 

Owing to the scaling property mentioned below Eq.~\eq{etaj}, the quantum
corrections for a given billiard shape depend only on $\theta_R$
and~$\theta_D$. These parameters are proportional to the size of the system.
Thus, in a material with definite physical SO-coupling constants $\alpha_R$
and~$\alpha_D$, Fig.~\ref{antiloc_RD_fig} (curves with constant $\theta_R/
\theta_D$) predicts the suppression of WA with the decreasing size of a chaotic
quantum dot (also observed in experiments~\cite{zumb02}). Preserving the shape
of the cavity means, in particular, that $\msf P_c /(w + w^\prime)$ is kept
constant when the dot size is changed. Alternatively, one can fix $w +
w^\prime$: this leads to a disproportionate decrease of $L_{\text{esc}}$ when
the size is reduced, thereby suppressing WA even stronger. 

In experiments, for fixed~$\theta_D$, $\theta_R$~can be tuned through an
additional back-gate potential. The dotted curve in Fig.~\ref{antiloc_RD_fig}
shows the result of such a parameter variation, including the associated
WA-WL-WA crossover.

\begin{figure}[tbp]
  \vspace*{.5cm}
  \includegraphics[width=.8 \linewidth, angle=0]{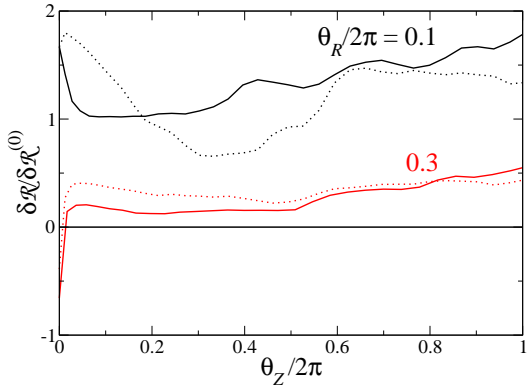}
  \caption{The relative quantum correction to reflection (or transmission)
  coefficients as a function of the Zeeman-coupling strength~$\theta_Z$ for the
  chaotic DS billiard with $\msf P_c /(w + w^\prime) = 90$. The in-plane
  magnetic field is parallel (solid curves) or perpendicular (dotted curves) to
  the leads' direction (see Fig.~\ref{bill_fig}). Two Rashba-coupling strengths
  $\theta_R/2 \pi = 0.1, \, 0.3$ (shown in the graph) are considered. The
  Dresselhaus interaction is absent. Each data point represents the average
  over 50,000 orbits. 
  \label{antiloc_Z_fig}}
\end{figure}

The effect of Zeeman interaction in the chaotic DS~billiard is considered in
Fig.~\ref{antiloc_Z_fig}. The interaction strength is measured by the
spin-precession angle per bounce~$\theta_Z = L_b |\mathbf C_Z| / |\mathbf v|$,
where $\mathbf C_Z$ is the Zeeman contribution to the effective field~$\mathbf
C$. We observe a complete suppression of WA even by weak Zeeman coupling. In
general, the $\theta_Z$~dependence of $\delta \mc R/ \delta \mc R^{(0)}$ is
very complex. It shows strong anisotropy with respect to the direction of the
applied in-plane magnetic field, especially, for weak SO interaction. Note that
the symmetry of the spin relaxation under the $\Lambda_R \leftrightarrow
\Lambda_D$ exchange (Sec.~\ref{efmf}) is lifted, unless $\mathbf C_Z$ lies in
the $(110)$ plane. 

\subsection{Comparison with full quantum mechanics}

\begin{figure}[tbp]
  \vspace*{.5cm}
  \includegraphics[width=.8 \linewidth, angle=0]{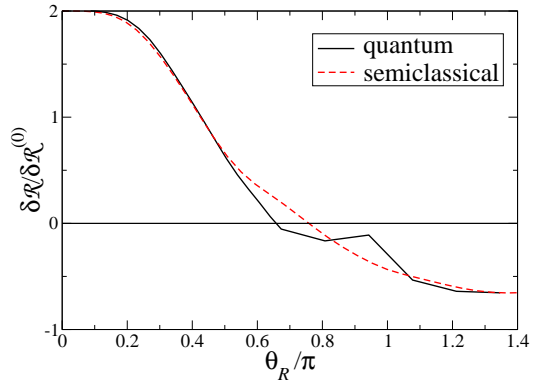}
  \caption{Relative quantum correction to reflection coefficient \emph{vs.}\
  Rashba-interaction strength~$\theta_R$ ($\theta_D = 0$) in the DS~billiard
  with $\msf P_c /(w + w^\prime) = 30$. The quantum reflection was computed for
  the energy corresponding to $N = N^\prime = 6$ open channels in the leads.
  The relative quantum correction (solid curve) was obtained after estimating
  the classical reflection $\mc R_{\text{cl}} \simeq 6.49$ as described in the
  text. In the semiclassical case (dashed curve) each data point represents the
  average over 50,000 orbits. 
  \label{antiloc_num_fig}}
\end{figure}

\begin{figure}[tbp]
  \vspace*{.5cm}
  \includegraphics[width=.95 \linewidth, angle=0]{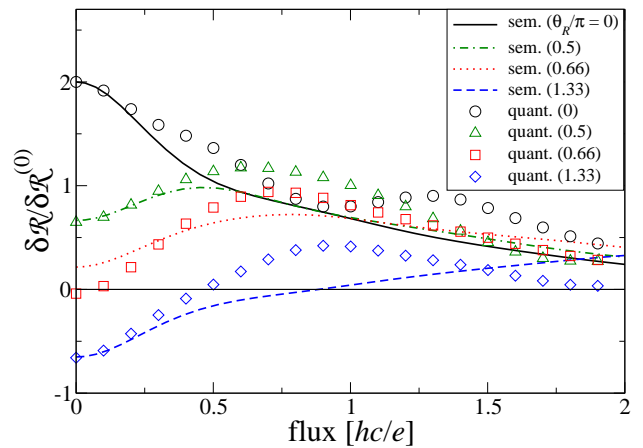}
  \caption{The relative quantum correction to reflection coefficient
  \emph{vs.}\ magnetic flux in the DS~billiard with $\msf P_c /(w + w^\prime) =
  30$. The quantum reflection was computed for the energy corresponding to $N =
  N^\prime = 6$ open channels in the leads. In the semiclassical case each data
  point represents the average over 50,000 orbits. The Rashba interaction was
  kept constant at $\theta_R / \pi = 0$ (solid curve: semiclassical~/~circles:
  quantum); $0.5$ (dash-dotted curve~/~triangles); $0.66$ (dotted
  curve~/~squares); $1.33$ (dashed curve~/~diamonds). The Dresselhaus and
  Zeeman couplings were not included.   
  \label{antiloc_flux_fig}}
\end{figure}

The semiclassical results are compared with quantum calculations in
Figs.~\ref{antiloc_num_fig} and~\ref{antiloc_flux_fig}. The complete quantum
transmission and reflection amplitudes $t_{n \sigma^\prime, m \sigma}$ and
$r_{n \sigma^\prime, m \sigma}$ are computed numerically by using a recursive
Green-function technique (see Ref.~\onlinecite{frus04b} and references therein
for details). This approach is based on a tight-binding model arising from the
real-space discretization of the \emph{spin-dependent} Schr\"odinger equation
in a two-dimensional geometry. Here, the standard on-site and hopping energies
present in a spinless tight-binding model are replaced by $2 \times 2$ matrices
accounting for the spin degree of freedom.\footnote{The Zeeman contribution due
to spin coupling to an external (possibly inhomogeneous) magnetic field appears
only in the on-site terms since it does not involve spacial derivatives. On the
contrary, the SO~coupling depending on the momentum $\mathbf p$ contributes
mainly to the hopping terms.} In combination with a recursive algorithm, we
obtain the Green function describing spin-dependent transport between the leads
by solving the (implicit) Dyson equation. The amplitudes $t_{n \sigma^\prime, m
\sigma}$ and $r_{n \sigma^\prime, m \sigma}$ are evaluated by projecting the
Green function onto a complete set of asymptotic states defining the
incoming/outgoing spin and orbital channels in the leads. The method has the
advantage of large flexibility: the quantum amplitudes can be equally
calculated for geometries of arbitrary shape. Moreover, disorder, magnetic
fields and spin-dependent interactions can be easily introduced and modified. 

The dependence of $\delta \mc R/ \delta \mc R^{(0)}$ on the Rashba-coupling
strength in the DS~billiard is presented in Fig.~\ref{antiloc_num_fig}.
Compared to Fig.~\ref{antiloc_cha_int_fig}, here we choose a smaller ratio
$\msf P_c /(w + w^\prime)$ in order to reduce the computing time in the quantum
case. There is a good overall agreement between the two curves, especially,
above the WL-WA transitional region. Since full quantum calculations give only
the total reflection, one is faced with the problem of separating the quantum
corrections from the classical part. Instead of evaluating $\mc R_{\text{cl}}$
directly, we estimate it by requiring that the ratio $\delta \mc R (\theta_R =
0) / \delta \mc R (\theta_R \gg 1)$ is the same in the semiclassical and
quantum consideration.

We used the estimated value of~$\mc R_{\text{cl}}$ for calculating the quantum
corrections in the presence of magnetic flux~$B_z \msf A_c/\Phi_0$
(Fig.~\ref{antiloc_flux_fig}). In this graph the quantum results exhibit
additional oscillations around the semiclassical curves, both with and without
the SO~interaction. The observed deviations from the semiclassics might be due
to the fact that, for the energy range considered in the quantum calculations,
the bending of classical trajectories caused by a Lorentz force limits the
applicability of the theory in Sec.~\ref{sdqc}. By the same token, the change
of the trajectories resulting from the SO~interaction might be responsible for
the discrepancy in Fig.~\ref{antiloc_num_fig}. Nevertheless, the semiclassical
techniques prove to be an effective tool for making reasonable predictions with
a minimum of computational power.\footnote{It took minutes to produce the
semiclassical curves in Figs.~\ref{antiloc_num_fig} and~\ref{antiloc_flux_fig},
and a couple of weeks for their quantum counterparts.}

The maximum in $\delta \mc R (B_z)/ \delta \mc R^{(0)}$, clearly noticeable at
the intermediate SO~interaction (Fig.~\ref{antiloc_flux_fig}), can be
attributed to the cancellation between the external flux and the flux of the
SO~vector potential~\eq{A1}\cite{crem03}. Equivalently, the integration~\eq{al}
with the modulation factor~\eq{Mphich} yields [cf. Eq.~\eq{wl}]
\eqn{
  \frac {\delta \mc R (\mathbf B)} {\delta \mc R^{(0)}} \simeq \sum_{\pm} \frac
  1 {1 + ( \wtilde B \pm \wtilde \Lambda^{-2})^2 L_{\text{esc}} / L_b}.
}
Thus, the maxima are located at $\wtilde B \approx \pm \wtilde \Lambda^{-2}$.
Note that the above expression, being positive, does not describe the WA, which
appears due to the $\delta \bs \eta^\parallel$~spin rotation (see
Sec.~\ref{lssm}).

\section{Summary and conclusions}

The semiclassical Landauer formula with spin expresses quantum corrections to
transmission and reflection in ballistic quantum dots as a sum over pairs of
classical trajectories or their loop parts, related by time reversal. The
effect of the spin-orbit and Zeeman interaction, as well as the Aharonov-Bohm
phase, can be taken into account by averaging of a modulation factor over the
ensemble of closed orbits. The change of classical trajectories by the
spin-orbit interaction and the Lorentz force can be neglected, provided the
energy of the particle is sufficiently high. With the increasing spin-orbit
coupling strength the quantum corrections may reverse their signs, and
weak localization will become weak antilocalization. Whether such a
transition takes place, depends on the interplay between the average spin
relaxation and the dwell time in the quantum dot. 

The spin relaxation, particularly for Rashba and Dresselhaus interaction, is
very sensitive to the character of classical dynamics, which is determined by
the shape of the boundary. While in chaotic geometries spin eventually
completely relaxes, in integrable systems (except for the circle billiard) the
relaxation is strongly suppressed. Consequently, at a given spin-orbit coupling
strength, a chaotic cavity can be in the weak-antilocalization regime, whereas
in a corresponding integrable cavity weak localization will take place. Weak
antilocalization is suppressed by an external magnetic field via the Zeeman
interaction and the Aharonov-Bohm phase. The size reduction of the quantum dot
works against the spin relaxation and the antilocalization, as well. Rashba and
Dresselhaus interactions compensate the effects of each other on
spin relaxation and transport.

The degree of spin relaxation for closed orbits, which controls the quantum
corrections to transport coefficients, is reduced, compared to open
trajectories of the the same length. This brings about an important difference
in the semiclassical treatment of systems with and without spin-orbit coupling:
in the latter case, the modulation factor due to magnetic flux can be averaged
over arbitrary trajectories. In the limit of slow spin dynamics, the
restriction to closed orbits is lifted, if the leading-order Rashba/Dresselhaus
terms are removed from the Hamiltonian by a gauge transformation. 

The semiclassical results for a chaotic billiard with moderate spin-orbit
interaction and weak magnetic field show a good agreement with the quantum
calculations. To account for the deviations at larger values, it may be
necessary to include the distortion of classical trajectories in the
semiclassical analysis. Possible extensions of the theory could be based on the
extended-phase-space approach to spin-orbit coupling~\cite{plet02, plet03}.

\begin{acknowledgments}

We thank L.~E.~Golub for helpful discussions. The work was supported by the
Deutsche Forschungsgemeinschaft through the Research group FOR~370 and the
Research school GRK~638 (OZ and KR) and by the EU Spintronics Research Training
Network~(DF).

\end{acknowledgments}

\appendix

\section{Spin propagator}
\label{sp}

Let us consider a spin, precessing in a time-dependent magnetic field~$\mathbf
C (t)$, with the Hamiltonian $\hat H (t)= \hbar \hat{\mathbf s} \cdot \mathbf C
(t)$. The path-integral~\eq{spinpi} for the corresponding propagator $K
(\zeta_2, \zeta_1; t)$ in the coherent-state representation
yields~\cite{koch95}
\eqn{
  K (\zeta_2, \zeta_1; t) = \frac {[a^*(t) - b^*(t) \zeta_1 + b(t) \zeta^*_2 +
  a(t) \zeta^*_2  \zeta_1 ]^{2s}} {(1 + |\zeta_2|^2)^s (1 + |\zeta_1|^2)^s},
\label{scsprop}
}
where the coefficients $a(t)$ and $b(t)$ are found from the differential
equation 
\eqn{
  \frac {dW (t)} {dt} = - \frac i 2\, \bs \sigma \cdot \mathbf C (t)\, W (t),
  \quad W(0) = \1
\label{Weqn}
}
for the matrix 
\eqn{
  W(t) \equiv \left(
  \begin{array}{cc}
    a(t) & b(t) \\
    - b^*(t) & a^*(t)
  \end{array}
  \right) \in \text{SU(2)}.
\label{W}
} 
In Eq.~\eq{Weqn} $\bs \sigma$ is the vector of Pauli matrices. As any SU(2)
matrix, $W(t)$ can be represented in the form: 
\aln{
  W(t) &= \1 \cos \frac {\eta (t)} 2 - i \bs \sigma \cdot \mathbf m (t) \, 
  \sin \frac {\eta (t)} 2 \notag \\
  &= e^{- i \bs \sigma \cdot \bs \eta (t)/ 2 },
\label{Wrot}
}
where $\mathbf m (t) \equiv \bs \eta (t)/ \eta (t)$ is a unit vector. The
spin~$s$ time-evolution operator (propagator) $\hat K(t)$ that belongs to the
$(2s + 1)$-dimensional irreducible representation of SU(2), satisfies the
equation 
\eqn{
  \frac {d \hat K} {dt} = - \frac i \hbar \hat H (t) \hat K = - i \hat{\mathbf
  s} \cdot \mathbf C (t)\, \hat K, \quad \hat K(0) = \hat \1,
\label{spinprop}
}
yielding 
\eqn{
  \hat K(t) = e^{ - i \hat{\mathbf s} \cdot \bs \eta (t)}.  
\label{spprgen}
}
Thus, $\hat K(t)$, applied to the spin states at time $t=0$, rotates them by
the angle~$\eta (t)$ about the axis~$\mathbf m (t)$. The coherent-state
propagator~\eq{scsprop} is the matrix element of~$\hat K(t)$:
\eqn{
  K (\zeta_2, \zeta_1; t) = \langle \zeta_2 | \hat K(t) | \zeta_1 \rangle =
  \langle \zeta_2 | e^{- i\mspace{2mu} \hat{\mathbf s} \cdot \bs  \eta(t)} |
  \zeta_1 \rangle.
\label{scsmatr}
}
Note that, setting $t=0$ (i.e., $a = 1, b = 0$) in Eq.~\eq{scsprop}, we obtain
an explicit expression for the scalar product $\langle \zeta_2 | \zeta_1
\rangle$ of two coherent states.

It is instructive to verify Eq.~\eq{scsmatr} in the case when the $z$~axis is
chosen along~$\mathbf m (t)$. Then $\hat K(t)$ is diagonal in the  $|\sigma
\rangle$~basis (Sec.~\ref{sost}): $\langle \sigma | \hat K | \sigma^\prime
\rangle = e^{- i \sigma \eta}\, \delta_{\sigma \sigma^\prime}$. Projecting
the spin coherent states~\eq{scs} onto the basis states,
\eqn{
  \langle \sigma | \zeta \rangle = \frac {\zeta^{s + \sigma}} {(1 + | \zeta
  |^2)^s}  \sqrt{ \frac {(2s)!} {(s + \sigma)!\, (s - \sigma)!} },
}
we find from Eq.~\eq{scsmatr}
\aln{
  &K (\zeta_2, \zeta_1) \notag \\
  &= \sum_{\sigma = -s}^s \frac {(\zeta_2^* \zeta_1)^{s + \sigma}\, (e^{-i
  \eta})^\sigma} {(1 + |\zeta_2|^2)^s\, (1 + |\zeta_1|^2)^s}\, \frac {(2s)!}
  {(s + \sigma)!\, (s - \sigma)!} \notag \\
  &= \frac {(e^{i \eta/2} + e^{- i \eta/2}\, \zeta_2^* \zeta_1)^{2s}}
  {(1 + |\zeta_2|^2)^s\, (1 + |\zeta_1|^2)^s}.
}
This result agrees with Eq.~\eq{scsprop} for $a(t) = e^{- i \eta(t)/2}$ and
$b(t) = 0$.

\section{Loop contribution}
\label{lc}

In Ref.~\onlinecite{rich02} the loop correction was derived for a hyperbolic
system with a single Lyapunov exponent~$\lambda$. Here we extend this
derivation to the case with SO~interaction, or, indeed, any interaction that
does not change the classical trajectories in the leading semiclassical
order.  

The action difference $\Delta \mc S (\varepsilon_{\gamma, \gamma^\prime})
\simeq M v^2 \varepsilon_{\gamma, \gamma^\prime}^2 / 2 \lambda$ for a loop pair
of trajectories $\gamma$ and $\gamma^\prime$ (in the absence of magnetic field)
depends on the crossing angle~$\varepsilon_{\gamma, \gamma^\prime}$, as well as
the velocity~$v$ and the mass~$M$. Any long orbit has many self-crossings, each
corresponding to a loop.  For an orbit of length~$L$, the average number of
loops of length between $l$ and $l + dl$ and crossing angle between
$\varepsilon$ and $\varepsilon + d \varepsilon$~is
\eqn{
  P_{\text{loop}} (l, \varepsilon; L)\, dl\, d \varepsilon \simeq 
  \frac {(L - l) \sin \varepsilon} {\pi \msf A_c}\, dl\, d \varepsilon.
\label{loopnum}
}
The loop length lies in the interval $l_{\text{min}} (\varepsilon) < l < L$,
where $l_{\text{min}} (\varepsilon) \sim -(2v / \lambda) \ln (\varepsilon/
\text{const})$ is the minimal possible loop length. This length is of order of
the size of the crossing region where the intersecting segments of the
trajectory are interrelated by the linearized dynamics and, therefore, cannot
form a closed loop. 

The loop correction to the channel-resolved transmission
coefficient~$|t_{nm}|_{\text{loop}}^2$ can be expressed as a sum over all
trajectories~$\gamma (\bar n, \bar m)$ weighted with the number of
loops~\eq{loopnum} per orbit:
\aln{
  |t_{nm}|_{\text{loop}}^2 =\, &2 \sum_{\substack{\text{loop} \\ \text{pairs}}}
  \mc A_\gamma \mc A_{\gamma^\prime}^* \cos \left[ \frac {\Delta \mc S
  (\varepsilon_{\gamma, \gamma^\prime})} \hbar \right]\, (\mc M
  \varphi)_{\gamma, \gamma^\prime} \notag \\
  \simeq\, &2 \sum_{\gamma (\bar n, \bar m)} |\mc A_\gamma|^2
  \int_0^\pi d \varepsilon \int_{l_{\text{min}} (\varepsilon)}^{L_\gamma} dl\, 
  P_{\text{loop}} (l, \varepsilon; L_\gamma) \notag \\
  &\times \cos \left[\frac {\Delta \mc S (\varepsilon)} \hbar \right]\, \wbar
  {\mc M \varphi}\, [l - l_{\text{min}} (\varepsilon)].
\label{loopint}
}
By $\wbar {\mc M \varphi}\, [l - l_{\text{min}} (\varepsilon)]$ we denote the
average of~$(\mc M \varphi)_{\gamma, \gamma^\prime}$ over all loops with the
effective length $l_e = l - l_{\text{min}} (\varepsilon)$. The contribution of
the crossing region to~$\wbar {\mc M \varphi}$ can be neglected, since, in the
semiclassical limit $\Delta \mc S (\varepsilon) / \hbar \gg 1$, only
small~$\varepsilon$ are important in the integral~\eq{loopint}. Moreover, only
the remaining part of the loop of length~$l_e$ obeys the statistical
assumptions under which the average~$\wbar {\mc M \varphi}$ is obtained.
Substituting $P_{\text{loop}} (l, \varepsilon; L_\gamma)$ from
Eq.~\eq{loopnum}, we compute the $l$~integral
\aln{
  &\int_{l_{\text{min}} (\varepsilon)}^{L_\gamma} dl\,
  P_{\text{loop}} (l, \varepsilon; L_\gamma)\, \wbar {\mc M \varphi}\, [l -
  l_{\text{min}} (\varepsilon)] \notag \\
  &\approx \frac {\sin \varepsilon} {\pi \msf A_c} \left[ \int_0^{L_\gamma}
  dl_e\, (L_\gamma - l_e)\, \wbar {\mc M \varphi} (l_e) \right. \notag \\
  &\phantom{\approx \frac {\sin \varepsilon} {\pi \msf A_c} \left[ \right.} 
  \left. {} - l_{\text{min}} (\varepsilon)
  \int_0^{L_\gamma} dl_e \, \wbar {\mc M \varphi} (l_e) \right]
\label{lint}
}
ignoring the terms of order~$[l_{\text{min}} (\varepsilon)]^2$. When
integrating over~$\varepsilon$ in Eq.~\eq{loopint}, we can approximate $\sin
\varepsilon \approx \varepsilon$ and extend the integration to infinity. Only
the term proportional to~$l_{\text{min}} (\varepsilon)$ in Eq.~\eq{lint}
survives the integration, and we obtain~\cite{sieb01}
\aln{
  &\int_0^\pi d \varepsilon\, \cos \left[\frac {\Delta \mc S (\varepsilon)}
  \hbar \right]\,  \frac {\sin \varepsilon} {\pi \msf A_c}\, l_{\text{min}}
  (\varepsilon)  \notag \\
  &\simeq [2 (N + N^\prime) L_{\text{esc}}]^{-1}.
\label{epsint}
}
Using the results of Eqs.~\eq{lint} and~\eq{epsint} in Eq.~\eq{loopint} and
applying the sum rule~\eq{sumrule}, we factorize the transmission coefficient
into a spin- and field-independent part and a length-averaged modulation
factor~$\langle \wbar{\mc M  \varphi}\, (\mathbf B) \rangle_L$:
\aln{
  &|t_{nm}|_{\text{loop}}^2 \notag \\
  &\simeq - (N + N^\prime)^{-2}\, L_{\text{esc}}^{-1} \int_0^\infty dL\, P_L
  (L) \int_0^L dl_e\, \wbar {\mc M \varphi} (l_e) \notag \\ 
  &= -(N + N^\prime)^{-2} \int_0^\infty dl_e\, P_L (l_e)\, \wbar {\mc M
  \varphi} (l_e).
}
After the summation over the channels we obtain the loop corrections~\eq{al}.

\section{Modulation factor for an~arbitrary spin}
\label{mfas}

To calculate the spin~$s$ modulation factor for a diagonal or loop pair of
trajectories,  we start from Eqs.\ \eq{Mdiag} and~\eq{Mloop}. Writing $\hat K$
in the exponential form~\eq{spprgen}, one finds
\eqn{
  \mc M = \text{Tr}\, (\hat K^2) = \text{Tr}\, (e^{-2i\, \hat{\mathbf s} \cdot
  \bs \eta}) = \sum_{\sigma = -s}^s e^{-2i \eta \sigma}.
}
Here, the last identity was obtained by choosing the $z$~axis in the direction
of~$\bs \eta$ and using the standard matrix form of~$\hat s_z$. It is
sometimes convenient to represent the result in terms of the Gegenbauer
polynomials~\cite{grad65}~$C_j^1$ on the sphere~$S^3$, which
generalize the Legendre polynomials on~$S^2$. For this purpose we transform
\aln{
  \sum_{\sigma = -s}^s e^{-i 2 \eta \sigma} &= \sum_{j = 0}^{2s} (-1)^j 
  \sum_{\sigma = -(2s - j)}^{2s - j} e^{-i  \eta \sigma} \notag \\
  &= \sum_{j = 0}^{2s} (-1)^{2s - j}\, \frac {\sin\, [(2j + 1)\, \eta / 2]} 
  {\sin\, (\eta / 2)}
}
and apply the property~\cite{grad65}
\eqn{
  \frac {\sin\, [(2j + 1)\, \eta / 2]}  {\sin\, (\eta / 2)} = C_{2j}^1
  \left(\cos \frac \eta 2 \right).
}
Eventually, we arrive at the expansion
\eqn{
  \mc M = \sum_{j = 0}^{2s} (-1)^{2s - j}\, C_{2j}^1 (\xi_4)
\label{Ms}
}
with $\xi_4 = \cos \frac \eta 2$ [see Eq.~\eq{xi}].

In the asymptotic limit $L \to \infty$ of long trajectories, when the spin
state is completely randomized, $\wbar{\mc M} \, (\infty)$ can be determined by
averaging of Eq.~\eq{Ms} over~$S^3$, which amounts to computing the scalar
product of $\mc M (\xi_4)$ and $C_0^1 (\xi_4) = 1$ on the sphere. Then the
orthogonality condition for the Gegenbauer polynomials~\cite{grad65},
\eqn{
  \frac 1 {2 \pi^2} \int_{S^3} d \bs \xi\, C_j^1 (\xi_4)\, C_{j^\prime}^1
  (\xi_4) = \delta_{j j^\prime},
\label{orth}
}
where $d \bs \xi$ is the surface element and $2 \pi^2$ is the surface ``area,''
yields $\wbar{\mc M} \, (\infty) = (-1)^{2s}$.

\section{Spin diffusion}
\label{sd}

The spin evolution in a random magnetic field is mapped onto a random walk on
the sphere~$S^3$ with initial point $\mathbf e_4 \equiv (0,0,0,1)$. In the
short-step limit $|\mathbf C|\, \tau /2 \ll \pi /2$, the probability
density~$P_{\text{diff}}\, (\bs \xi; t)$ to find the particle at point~$\bs
\xi$ at time~$t$ solves the diffusion equation~\cite{mond98}
\eqn{
  \left( \frac \pd {\pd t} - D \nabla^2 \right) P_{\text{diff}}\, (\bs \xi; t)
  = 0, \quad P_{\text{diff}}\, (\bs \xi; 0) = \delta (\bs \xi, \mathbf e_4),
} 
with the diffusion coefficient $D = |\mathbf C|^2\, \tau /24$. Here $\nabla^2$
is the Laplacian on~$S^3$, and $\delta (\bs \xi, \mathbf e_4)$ is the
$\delta$-function on~$S^3$ between two points. The probability is normalized by
$\int_{S^3} d \bs \xi\, P_{\text{diff}}\, (\bs \xi; t) = 1$, where $d \bs \xi$
is the surface element. The solution can be found in the the form of an
expansion in the Gegenbauer polynomials~\cite{grad65}~$C_j^1$:
\eqn{
  P_{\text{diff}}\, (\bs \xi; t) = \frac 1 {2 \pi^2} \sum_{j = 0}^\infty (j +
  1)\,  e^{-j (j + 2) D t}\, C_j^1 (\xi_4).
}

In order to determine the average modulation factor~$\wbar{\mc
M}_{\text{diff}}\, (t)$, we express $\mc M (\xi_4) = C_2^1 (\xi_4) - C_0^1
(\xi_4)$ from Eq.~\eq{M12}. Taking into account the orthogonality
condition~\eq{orth}, we compute the average for the
distribution~$P_{\text{diff}}\, (\bs \xi; t)$,
\eqn{
  \wbar{\mc M}_{\text{diff}}\, (t) = \int_{S^3} d \bs \xi\, P_{\text{diff}}\,
  (\bs \xi; t)\, \mc M (\xi_4) = 3\, e^{-8Dt} - 1, 
}
yielding Eq.~\eq{Mdiff}. This result can be generalized to an arbitrary
spin~$s$ if one employs the representation~\eq{Ms}:
\eqn{
  \wbar{\mc M}_{\text{diff}}\, (t) = \sum_{j = 0}^{2s} (-1)^{2s - j}\, (2j +
  1)\,  e^{-4j (j + 1) Dt}.
}
The average spin polarization is, according to Eq.~\eq{nz}, 
\eqn{
  \wbar{(n_z)}_{\text{diff}}\, (t) = 2 [ \wbar{\xi_3^{\, 2}} (t) +
  \wbar{\xi_4^{\, 2}} (t)] - 1.
} 
Using Eqs.\ \eq{M12} and~\eq{Mdiff}, we find 
\eqn{
  \wbar{\xi_4^{\, 2}} (t) = (3\,
  e^{- \frac 1 3 |\mathbf C|^2 \tau t} + 1)/4. 
}
The symmetry of the problem
provides $3\, \wbar{\xi_3^{\, 2}} (t) = 1 - \wbar{\xi_4^{\, 2}} (t)$, and
Eq.~\eq{ndiff} follows.

\section{Area contribution to the spin-rotation angle}
\label{ac}

\begin{figure}[tbp]
  \vspace*{.5cm}
  \includegraphics[width=.6 \linewidth, angle=0]{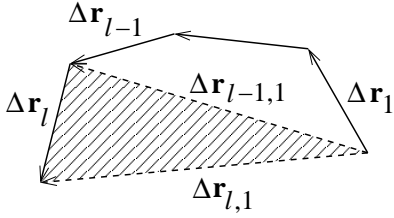}
  \caption{Definition of the enclosed area [Eq.~\eq{area}]. The shaded area is
  $(1/2)\, \Delta \mathbf r_{l-1,1} \times\Delta \mathbf r_l$. 
  \label{area_fig}}
\end{figure}

To derive Eq.~\eq{etaperp}, it is convenient to use the notation $\bs
\eta_{l,1} \equiv \bs \eta$ in Eq.~\eq{Wtr}. When $l=2$, the BCH
formula~\eq{BCH} applied to the product of the two segment propagators yields
\eqn{
  \bs \eta_{2,1} = \bs \eta_1 + \bs \eta_2 - \frac 1 2 \, \bs \eta_1 \times \bs
  \eta_2 + \mc O (\Lambda_{X, Y}^{\, -3}). 
\label{eta21}
}
For an arbitrary~$l$, this result generalizes to
\eqn{
  \bs \eta_{l,1} = \bs \eta_{l,1}^{(0)} - \frac 1 2 \sum_{j=2}^l \bs
  \eta_{j-1,1}^{(0)} \times \bs \eta_j + \mc O (\Lambda_{X, Y}^{\, -3}),
\label{etal1}
}
where $\bs \eta_{l,1}^{(0)} = \sum_{j=1}^l \bs \eta_j$. The above expression
can be proven by induction in~$l$. Namely, we assume its validity for $l - 1$
and compute~$\bs \eta_{l,1}$ from
\eqn{
  e^{- i \bs \sigma \cdot \bs \eta_{l,1} / 2} = e^{- i \bs \sigma \cdot \bs
  \eta_l / 2}\, e^{- i \bs \sigma \cdot \bs \eta_{l-1,1} / 2},
}
as in Eq.~\eq{eta21}. Equation~\eq{etaperp} will follow from Eq.~\eq{etal1} if
the rotation angles are expressed in terms of displacements using
Eqs.~\eq{etaj} and~\eq{eta0}. The enclosed area is defined by 
\eqn{
  \msf A\, \mathbf e_z = \frac 1 2 \sum_{j=2}^l \Delta \mathbf r_{j-1,1} \times
  \Delta \mathbf r_j,
\label{area}
}
where $\Delta \mathbf r_{j,1} = \sum_{i=1}^j \Delta \mathbf r_i$
(Fig.~\ref{area_fig}).

\bibliography{antiloc_long_arxiv}

\end{document}